\documentclass[epj]{svjour}
\usepackage{epsfig}

\begin{document}

\title{Proton-proton scattering above 3 GeV/c}
\author{A.~Sibirtsev\inst{1,2,3}, J.~Haidenbauer\inst{3,4},  
H.-W. Hammer\inst{1}, S.~Krewald\inst{3,4} and
U.-G.~Mei{\ss}ner\inst{1,3,4}
}                     


\institute{
Helmholtz-Institut f\"ur Strahlen- und Kernphysik (Theorie)
und Bethe Center for Theoretical Physics,
Universit\"at Bonn, D-53115 Bonn, Germany  \and
Excited Baryon Analysis Center (EBAC), Thomas Jefferson National
Accelerator
Facility, Newport News, Virginia 23606, USA
\and Institut f\"ur Kernphysik and J\"ulich Center for Hadron Physics,
Forschungszentrum J\"ulich, D-52425 J\"ulich, Germany
\and Institute for Advanced Simulation,
Forschungszentrum J\"ulich, D-52425 J\"ulich, Germany
}

\date{Received: date / Revised version: date}

\abstract{A large set of data on proton-proton differential cross sections,
analyzing powers and the double polarization parameter $A_{NN}$ is analyzed
employing the Regge formalism. We find that the data available at
proton beam momenta from 3 GeV/c to 50 GeV/c exhibit features that 
are very well in line with the general characteristics of Regge
phenomenology and can be described with a model that includes the
$\rho$, $\omega$, $f_2$, and $a_2$ trajectories and single Pomeron
exchange. 
Additional data, specifically for spin-dependent observables at
forward angles, would be very helpful for testing and refining our
Regge model. 
} 

\authorrunning{A. Sibirtsev {\it et al.}}

\maketitle
\section{Introduction}
Nucleon-nucleon ($NN$) elastic scattering is {\it the} primary process to 
understand nuclear forces and to construct theoretical models of the 
strong interaction. While relativistic meson-exchange models
\cite{Kloet80,Kloet81,Dubach82,Lee84,Faassen84,Faassen86,Haar87,Machleidt87,HHJ,Machleidt01,Eyser04,Pricking07} 
work reasonably well at beam momenta below 2~GeV/c,
say, the situation is quite different
at higher proton momenta. The typical problems are summarized in
Refs.~\cite{Eyser04,Machleidt01a}. First, the predicted $NN$ elastic
cross sections are too large and increase with momentum, while
experimentally they decrease. Second, the predicted analyzing powers are
too large while for other spin observables even the sign differs from that
observed in experiments. The first problem is caused by the vector-meson
exchange in the standard meson-exchange models. It was claimed~\cite{Machleidt01a}
that, because of the first problem, Regge theory was invented since it generates
the appropriate energy dependence of the scattering amplitude. The second
problem, however, persists even when the contribution from vector mesons is 
taken off.
Thus, the momentum range from 2~GeV/c to 10~GeV/c offers precise and still
unexplained data, waiting for being described by adequate models. 

During the last years some progress was achieved in the analysis of
$NN$ scattering below $\simeq$ 4~GeV/c within an approach based 
on relativistic optical potentials~\cite{Geramb98,Neudachin91,Funk01,Knyr06}.
Other recent activities concern the application of the Regge formalism 
\cite{Buttimore99,Kochelev00,Kharzeev00,Donnachie01,Donnachie04} in
the analysis of proton-proton ($pp$) scattering at very high momenta, say 
above 100~GeV/c. Here the most interesting finding is that there are possibly
contributions from new Pomeron trajectories \cite{Donnachie04,Cudell05} 
that are insignificant at low energies.

The most recent extensive measurements of $pp$ scattering were done
by the EDDA Collaboration at COSY~\cite{Altmeier00,Altmeier05} and at
SATURNE in 
Saclay~\cite{Perrot87,Allgover99,Allgover99a,Allgower00,Allgower01}. These
experiments provide a wealth of precise data on differential cross sections
and polarizations up to beam momenta of about 3.8~GeV/c. Unfortunately, these
measurements do not cover forward angles below $\theta_{c.m.}\simeq$ 30$^\circ$.
A further improvement of the high-quality data base on $pp$ scattering
is expected from the ANKE Collaboration at COSY. Very recently it was
proposed~\cite{Chiladze09} to measure $pp$ elastic scattering at beam
momenta from 2.4~GeV/c to 3.6~GeV/c and at c.m. angles from 10$^\circ$ to 30$^\circ$.
The experiment aims to obtain precision data for differential
cross sections and analyzing powers. The experiment could be extended in the
future to measure double polarization observables. 

The findings in Refs.~\cite{Eyser04,Machleidt01a} 
show that traditional nuclear models of the $NN$ interaction such as
meson-exchange potentials cannot be easily extended to the energies 
covered by the COSY-EDDA and SATURNE experiments. At the same time that region 
is much too low for perturbative QCD to be applied. 
Thus, in this paper we propose to use a phenomenological Regge approach 
to study the $NN$ interaction at those COSY-EDDA and SATURNE energies. 

In the present paper we consider the $\rho$, $\omega$, $f_2$, 
and $a_2$ Regge exchanges, supplemented by the contribution of a single 
Pomeron Regge pole. We expect that those conventional Regge trajectories
represent the proper dynamical degrees of freedom for describing $NN$ scattering 
in the region from the COSY-EDDA and SATURNE energies up to beam momenta of 50 to 100 GeV/c.  
Indeed at the latter momenta the $pp$ and antiproton-proton ($\bar pp$) 
cross sections are already pretty close to each other, cf. the review section 
of the PDG~\cite{PDG}, which is a clear indication that the contributions 
from the mesonic Regge trajectories have become practically 
negligible and $pp$ (as well as $\bar pp$) scattering is dominated by 
Pomeron exchange alone. In fact, as already said above, at such high 
momenta additional dynamical mechanisms become more and more important which 
do not play a role for energies down to the COSY-EDDA and SATURNE regime. 

As will be shown in this paper, with contributions from the $\rho$, $\omega$, 
$f_2$, and $a_2$ Regge exchanges and a single Pomeron Regge pole
we are able to describe the
$pp$, $\bar pp$, $pn$, $\bar pn$ total cross sections,
but also the ratio of the real-to-imaginary parts of the forward amplitudes 
for $pp$ and $\bar pp$, over a large energy range up to beam momenta of 
around 100 GeV/c. 
Next, we use the data on angular distributions and polarizations for $pp$
to determine the details of the Regge exchanges. Obviously, it is an open
question down to which energies such a phenomenological Regge approach
can provide a quantitative description of the available data. We find
that the phenomenology allows to achieve a useful representation of the 
data down to 3~GeV/c, in line with analogous investigations for pion-nucleon 
scattering~\cite{Sibirtsev07,Sibirtsev09a,Huang09,Sibirtsev09,Huang09a}. 
The Regge region therefore turns out to have a significant overlap with 
the kinematical region covered by the COSY-EDDA and SATURNE $pp$ experiments. 

The paper is organized as follows: In sect.~2, we introduce the helicity
amplitudes for $pp$ scattering. In sect.~3 we describe the details of the
fitting procedure.
Sect.~4 provides a comparison of the results of our Regge model with data
on differential cross sections and polarizations. In sect.~5
a comparison between the Regge calculation and results based on the
partial wave analysis (PWA) of the George-Washington Group (GWU)
is made for some beam momenta. The paper ends with a summary. 
Detailed information on the data sets used in our analysis are
summarized in an Appendix.

\section{Helicity amplitudes}
In the past Regge analyses of $NN$ 
scattering at beam momenta from 3 GeV/c upwards were presented in
Refs.~\cite{Arbab67,Rarita68}. Unfortunately, the proposed formalism 
cannot be simply taken over for the study of the new data. In
Ref.~\cite{Arbab67} the amplitudes are expanded for very low
momentum transfer squared or for very forward angles. The analysis of
Ref.~\cite{Rarita68} is fairly complete but, strictly speaking, is not 
based on a 
covariant formulation. For instance the Regge propagator is expressed not
in terms of Mandelstam invariants like the squared invariant
collision energy $s$, as usually done, but in terms of the laboratory energy. 
This is in conflict with the energy dependence of the total $pp$ 
cross section, where there is strong evidence ~\cite{Cudell02,Pelaez06} that
its $s$-dependence is driven by the leading Regge trajectories. 

It is convenient to expand the $NN$ scattering matrix $\phi$
in terms of the helicity amplitudes proposed by Jacob and
Wick~\cite{Jacob59}: 
\begin{eqnarray}
 \phi_a = \langle \lambda_1^\prime \lambda_2^\prime |\phi| \lambda_1
\lambda_2 \rangle,
\end{eqnarray}
Here $\lambda_1 (\lambda_1^\prime)$ and $\lambda_2 (\lambda_2^\prime)$
are the $s$-channel projections of the spin of the initial (final)
protons, respectively. The index $a$ labels all possible
combinations of helicities for the transition between initial and final
states. Taking into account parity conservation and time reversal
invariance~\cite{Bystricky78} the full set of the covariant
$s$-channel helicity amplitudes is given by~\cite{Goldberger60}
\begin{eqnarray}
\phi_1 &=&\langle {+\frac{1}{2}}\, {+\frac{1}{2}} |\phi| {+\frac{1}{2}}\,
{+\frac{1}{2}} \rangle {=}  \langle {-\frac{1}{2}}\, {-\frac{1}{2}}   
|\phi| {-\frac{1}{2}}\, {-\frac{1}{2}} \rangle,  \nonumber \\
\phi_2 &=& \langle {+\frac{1}{2}}\, {+\frac{1}{2}} |\phi| {-\frac{1}{2}}\,
{-\frac{1}{2}} \rangle{=}  \langle {-\frac{1}{2}}\, {-\frac{1}{2}} |\phi|
{+\frac{1}{2}}\,{+\frac{1}{2}} \rangle, \nonumber \\
\phi_3 &=& \langle {+\frac{1}{2}}\, {-\frac{1}{2}} |\phi| {+\frac{1}{2}}\,
{-\frac{1}{2}} \rangle {=} \langle {-\frac{1}{2}}\, {+\frac{1}{2}} |\phi|
{-\frac{1}{2}}\, {+\frac{1}{2}} \rangle, \nonumber \\
\phi_4 &=& \langle {+\frac{1}{2}}\, {-\frac{1}{2}} |\phi| {-\frac{1}{2}}\,
{+\frac{1}{2}} \rangle {=} \langle {-\frac{1}{2}}\, {+\frac{1}{2}} |\phi|
{+\frac{1}{2}}\,{-\frac{1}{2}} \rangle, \nonumber \\
\phi_5 &=&  \langle {+\frac{1}{2}}\, {+\frac{1}{2}} |\phi| {+\frac{1}{2}}\,
{-\frac{1}{2}} \rangle {=} \langle {-\frac{1}{2}}\, {+\frac{1}{2}} |\phi|
{-\frac{1}{2}}\,{-\frac{1}{2}} \rangle\phantom{,} \nonumber \\
&=& \langle {-\frac{1}{2}}\, {-\frac{1}{2}} |\phi| {+\frac{1}{2}}\,
{-\frac{1}{2}} \rangle {=} \langle {-\frac{1}{2}}\, {+\frac{1}{2}} |\phi|
{+\frac{1}{2}}\,{+\frac{1}{2}} \rangle\phantom{,} \nonumber \\
&=& -\langle {-\frac{1}{2}}\, {-\frac{1}{2}} |\phi| {-\frac{1}{2}}\,
{+\frac{1}{2}} \rangle {=}-\langle {+\frac{1}{2}}\, {-\frac{1}{2}} |\phi|
{+\frac{1}{2}}\,{+\frac{1}{2}} \rangle  \nonumber \\
&=& -\langle {+\frac{1}{2}}\, {+\frac{1}{2}} |\phi| {-\frac{1}{2}}\,
{+\frac{1}{2}} \rangle {=} -\langle {+\frac{1}{2}}\, {-\frac{1}{2}} |\phi|
{-\frac{1}{2}}\,{-\frac{1}{2}} \rangle.
\end{eqnarray}
Therefore, elastic scattering of two identical particles with 
spin-$1/2$ can be
completely described by five independent amplitudes. The helicity
amplitudes $\phi_1$ and
$\phi_3$ given above correspond to helicity nonflip, $\phi_5$ is the
single helicity-flip amplitude, while $\phi_2$ and $\phi_4$ are double
helicity-flip amplitudes.

The helicity amplitudes depend on the invariant kinematic variables defined
as
\begin{eqnarray}
s&=&(p_1+p_2)^2=(p_1^\prime +p_2^\prime)^2, \nonumber\\
t&=&(p_1-p_1^\prime)^2=(p_2-p_2^\prime)^2,  \nonumber\\
u&=&(p_1-p_2^\prime)^2=(p_1^\prime-p_2)^2, \label{kinematics}
\end{eqnarray}
where $p_1$ ($p_1^\prime$) and $p_2$ ($p_2^\prime$) are the four-momenta
of the incident (final) protons, respectively, and
\begin{eqnarray}
s+t+u=4\,m_N^2,
\end{eqnarray}
with $m_N$ being the proton mass.

We consider $pp$ scattering via the exchange of Regge
trajectories defined by a certain set of allowed quantum numbers.
Within the Regge formalism the helicity amplitudes $\phi_a$ 
($a{=}1,...,5$) can be parameterized for each exchange trajectory $i$ by
\begin{eqnarray}
\phi_{ai}(s,t) &=& \pi \beta_{ai}(t) \frac{
\zeta_i(s,t)}{\Gamma\!\left(\alpha(t)\right)}, 
\label{amplitude}
\end{eqnarray}
where $\beta_{ai}$ is the product of the vertex functions and $\Gamma$
denotes the $\Gamma$-function that is introduced in order to suppress the
poles of the Regge propagator in the scattering region. The total 
helicity amplitudes are given by the corresponding sum over the contributing 
trajectories. Furthermore, $\zeta_i$ is the Regge propagator taken to be 
\begin{eqnarray}
\zeta_i(t,s)= \frac{1+{\cal S}_i \, {\rm
exp}\!\left[-i\, \pi \, \alpha_i(t)\right]}{{\rm sin}
[\pi \,  \alpha_i(t) ]} \left[\frac{s}{s_0}\right]^{\alpha(t)},
\label{rpropa}
\end{eqnarray}
with ${\cal S}_i$ being the signature of the trajectory and $s_0 = 1$~GeV$^2$ 
a scaling factor. The $i$-th 
Regge trajectory, $\alpha_i$, is considered as a linear function of $t$,
\begin{eqnarray}
\alpha_i(t) = \alpha^0_i + \alpha_i^\prime \,t,
\end{eqnarray}
with the slope $\alpha_i^\prime$ and the intercept $\alpha^0_i$ either being
determined by a fit to data or taken from the analysis of other
reactions. Note that the difference between
the power $\alpha(t)$ here and the power $\alpha(t){-}1$ given in our previous
publications~\cite{Sibirtsev07,Huang09,Sibirtsev09} on meson
photoproduction and pion charge-exchange is due to different normalizations,
{\it i. e.} different relations between the helicity amplitudes and observables.

The signature ${\cal S}_i$ of the exchange trajectory is
defined as follows. Both natural and unnatural parity
trajectories can be exchanged in the $t$-channel.
The naturalness $\cal N$ for natural (${\cal N}{=}{+1}$) and unnatural
(${\cal N}{=}{-1}$) parity exchanges is defined as
\begin{eqnarray}
{\cal N} &=& +1 \,\,\, \mathrm{if} \,\,\, P=(-1)^J, \nonumber \\
{\cal N} &=& -1 \,\,\, \mathrm{if} \,\,\, P=(-1)^{J+1},
\end{eqnarray}
where $P$ and $J$ are the parity and spin of the particle lying on the
Regge trajectory. The signature factor ${\cal S}{=}{\pm}1$
is then defined as~\cite{Irving,Collins2,Collins3}
\begin{eqnarray}
{\cal S} = P \times {\cal N} = (-1)^J.
\label{signat}
\end{eqnarray}

The structure of the functions $\beta_{ai}$ of Eq.~(\ref{amplitude})
is defined by the quantum numbers of the particles at the interaction
vertices, similar to the usual particle-exchange Feyman diagram. 
The general form of the functions $\beta_{ai}$ is given by~\cite{Irving}
\begin{eqnarray}
\beta_{ai}(t) {=}c_{ai} \,
F_{ai}(t)\left[\frac{-t}{4m_N^2}\right]^{\frac{
|\lambda_1^\prime-\lambda_1|}{2} }
\left[\frac{-t}{4m_N^2}\right]^{\frac{|\lambda_2^\prime-\lambda_2|}{2}},
\end{eqnarray}
where $c_{ai}$ is a constant and $F_{ai}$ is an overall form factor.
For the latter we take an exponential function so that in our case
the $\beta_{ai}$ are parameterized by 
\begin{eqnarray}
\beta_{1i}(t) &=& c_{1i} \, \exp(b_{1i}t), \nonumber \\
\beta_{2i}(t) &=& c_{2i} \, \exp(b_{2i}t) \, \frac{-t}{4m_N^2}, \nonumber \\
\beta_{3i}(t) &=& c_{3i} \, \exp(b_{3i}t), \nonumber \\
\beta_{4i}(t) &=& c_{4i} \, \exp(b_{4i}t) \, \frac{-t}{4m_N^2}, \nonumber \\
\beta_{5i}(t) &=& c_{5i} \,  \exp(b_{5i}t) \,
\left[\frac{-t}{4m_N^2}\right]^{1/2}.
\label{set}
\end{eqnarray}
Therefore, in general for each trajectory that contributes to 
$pp$ elastic scattering one should determine 10 free parameters, 
in case that the intercepts and slopes of the trajectories are known
from other sources.

In the asymptotic limit, i. e. for large $s$ and small $|t|$, the helicity
amplitudes exhibit the property~\cite{Sharp63,Itzykson63}
\begin{eqnarray}
\phi_{1i}=\phi_{3i} \hspace{3mm} {\rm and}  \hspace{3mm}
\phi_{2i}=-\phi_{4i} \ . 
\label{assymptotic}
\end{eqnarray}
This allows to reduce the number of free parameters to six for each
trajectory. However, that relation should be considered as a leading order
approximation and further improvements of the model might be achieved by
considering not only three but the complete set of five independent 
amplitudes as given by Eq.~(\ref{set}). 
Nevertheless, we try to keep the number of free
parameters as small as possible, because the amount of polarization
data, to which the calculations are very sensitive, is rather limited at
high energies. 

Let us now specify the trajectories that contribute to $pp$
scattering. At high energies hadron-hadron scattering is
dominated by Pomeron exchange. We consider here a single Pomeron Regge 
pole for which we adopt the parameters 
\cite{Donnachie84,Donnachie86,Donnachie92}
\begin{eqnarray}
 \alpha_P(t)=1.08+0.25 t \ .
\label{intercept}
\end{eqnarray}

\begin{figure}[t]
\vspace*{-6mm}
\centerline{\hspace*{5mm}\psfig{file=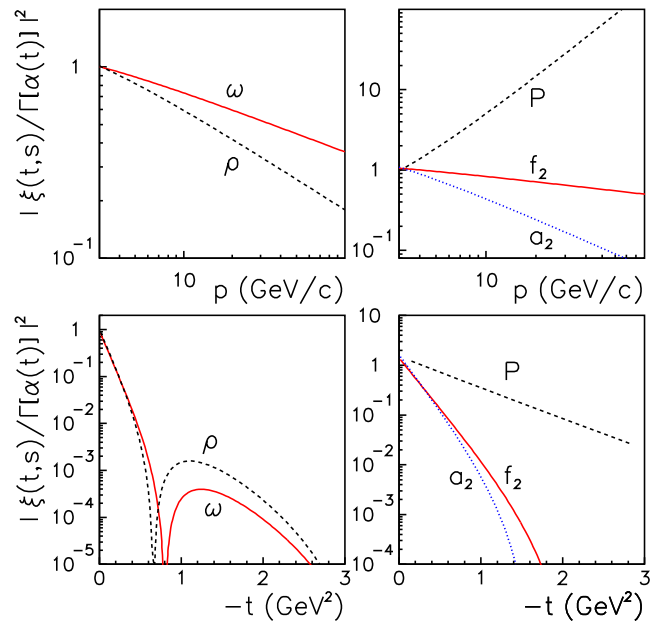,width=9.5cm}}
\vspace*{-5mm}
\caption{ \label{kinem2} Absolute square of the Regge propagator of
Eq.~(\ref{rpropa}) divided by the $\Gamma$-function as a function of the beam
momentum (upper panel) at fixed $t{=}{-}$1 GeV$^2$ and as a function of the 
four-momentum transfer squared (lower panel) at the beam momentum $p$=5 GeV/c. 
Results are presented for the $\rho$, $\omega$, $f_2$, $a_2$, and Pomeron (P)
trajectories as indicated in the plots.
}
\end{figure}

The $\rho$ and $\omega$ trajectories are taken from our global analysis
of pion charge-exchange~\cite{Huang09} and neutral pion
photoproduction~\cite{Sibirtsev09} data. Note that these
analyses include experimental results on differential cross sections and
on single as well as double polarizations. We take
\begin{eqnarray}
\alpha_\rho(t)=0.53+0.8 t \ ,  \nonumber \\
\alpha_\omega(t)=0.64+0.8 t \ .
\end{eqnarray}
For the $f_2$ and $a_2$ trajectories we adopt the values from Ref.
\cite{Anisovich00}: 
\begin{eqnarray}
\alpha_{f_2}(t)=0.71+0.83 t, \\
\alpha_{a_2}(t)=0.45+0.91 t,
\end{eqnarray}
These authors fixed the trajectories from a systematic analysis of the 
mesonic states in the Chew-Frautschi plot, given by the plane defined by 
the squared mass and the spin of the particles.

Pion exchange plays a significant role at low energies. 
Specifically, it constitutes the longest ranged part of the 
$pp$ interaction and it dominates at very small $-t$, i.e. close to the
pole of the pion propagator. 
Indeed, it was argued~\cite{Ericson95} that the pion pole term dominates
so strongly in forward direction, that one can actually
extract the $\pi{NN}$ coupling constant from the differential cross section
by an extrapolation to the pion pole. 
In the Regge model the pion trajectory is given
by~\cite{Anisovich00,Sibirtsev03}
\begin{eqnarray}
 \alpha_\pi(t)=0.7(t-m_\pi^2),
\end{eqnarray}
where $m_\pi$ is the pion mass. Hence the contribution from pion
exchange decreases much more rapidly with increasing beam momentum 
compared to those of the other contributions and, therefore, is difficult 
to determine. This problem in determining the contribution from the pion 
exchange trajectory is further intensified
by the lack of $pp$ scattering data in the forward direction 
at the lower energies considered in the present investigation. 
Thus, at this stage we do not include the pion trajectory.

It is instructive to look at the beam-momentum- and $t$ dependence for each
of the included trajectories. Fig.~\ref{kinem2} illustrates the absolute
squared Regge propagator of Eq.~(\ref{rpropa}) 
divided by the $\Gamma$-function,
{\it i. e.} $|\zeta_i(t,s)/\Gamma[\alpha_i(t)]|^2$ for different
trajectories. The upper panel of Fig.~\ref{kinem2} shows the dependence on 
the laboratory momentum at fixed $t{=}-$1~GeV$^2$. The results for
different trajectories are arbitrarily normalized at $p{=}3$~GeV/c. It is
clear that with increasing momentum, the Pomeron trajectory becomes
more and more significant. 

The lower panel of Fig.~\ref{kinem2} illustrates the $t$-dependence
of $|\zeta_i(t,s)/\Gamma[\alpha_i(t)]|^2$ at the beam momentum $p$=5 GeV/c. 
The zeros of the amplitudes due to the $\rho$ and $\omega$ trajectories are
close to each other, but not at exactly the same $t$. If one of the 
trajectories dominates the reaction we would expect that there is a 
minimum in the differential cross section at a certain value of $t$. 
Indeed the data on the $\gamma p {\to}\pi^0p$ and $\pi^- p{\to}\pi^0n$
reactions show
minima around ${-}t{=}0.5-0.6$~GeV$^2$. The minimum could be
shifted due to an interference between different trajectories. 
Moreover, it is possible that the minimum is not  observed at all, if 
the contributing trajectories play an equally important role, like in
the $\gamma{p}{\to}\pi^+n$ and $\gamma{n}{\to}\pi^-p$ reactions.

For scattering of identical particles the amplitudes 
$\phi_{a}$ $(a=1,...,5)$ obey certain symmetry relations. 
Specifically, for $pp$ scattering they read \cite{Bystricky78}:
$\phi_{1,2}(\theta_{c.m.}) = \phi_{1,2}(\pi - \theta_{c.m.})$,
$\phi_3(\theta_{c.m.}) = -\phi_4(\pi - \theta_{c.m.})$,
$\phi_5(\theta_{c.m.}) = -\phi_5(\pi - \theta_{c.m.})$.
We implement these symmetry relations via the substitution
$\phi_1(t) \rightarrow  \phi_1(t) + \phi_1(u)$, etc. 
\cite{Avilez}. This replacement has no influence on the
results for small $-t$. However, it modifies the
amplitudes for $t$ values corresponding to angles close to
$\theta_{c.m.} = 90^o$. In particular, then the analyzing power 
vanishes at $90^o$, as it must be the case for $pp$ scattering. 

Finally, the helicity amplitudes are normalized in such a way that the
differential cross section is given by
\begin{eqnarray}
 \frac{d\sigma}{dt}{=}\frac{1}{64 \pi q^2 s}\frac{1}{2}
[|\phi_1|^2{+}|\phi_2|^2{+} |\phi_3|^2{+}|\phi_4|^2{+}4|\phi_5|^2].
\end{eqnarray}
The analyzing power $A$ is then given by 
\begin{eqnarray}
A\propto -{\rm Im} \left[(\phi_{1}+\phi_{2}+\phi_{3}-\phi_{4})\,
\phi_{5}^\ast \right]~.
\end{eqnarray}
A complete overview of the relations between the helicity amplitudes 
and various observables can be found in Ref.~\cite{Bystricky78}. 

\section{Fit procedure}

In our global analysis of the $pp$ scattering data we include
five trajectories, namely the Regge amplitudes for 
$\omega$, $\rho$, $f_2$, $a_2$
exchanges and the Pomeron. In order to minimize the number of free
parameters we impose the asymptotic relations given in Eq.~(\ref{assymptotic}).
Therefore, there are 30 parameters that have to be determined in a fit to
data in order to fix the coupling constants and form factors of
Eq.~(\ref{set}). 
The parameters of the model are compiled in Table~\ref{param}.

\begin{table}[t]
\begin{center}
\caption{\label{param} 
Parameters 
of the model. The constants $c_{ai}$ 
are given in GeV${\cdot}\sqrt{mb}$ and the $b_{ai}$'s are given in
GeV$^{-2}$. 
}
\renewcommand{\arraystretch}{1.2}
\begin{tabular}{|c|c||r|r|r|}
\hline
  & $i$ & $c_{1i}$ & $c_{2i}$ & $c_{5i}$  \\
\hline
$P$      & 1 & -10.9$\pm$2.3 & 524$\pm$41 & -0.98$\pm$0.15  \\
$f_2$    & 2 & -19.0$\pm$1.3 &  -1699$\pm$170 & 322$\pm$4 \\
$\omega$ & 3 & 11.7$\pm$1.2 & 376$\pm$47 & -26.5$\pm$11.2  \\
$\rho$   & 4 & 2.6$\pm$0.3 & -216$\pm$60 & 50.3$\pm$7.1 \\
$a_2$   & 5 & -2.8$\pm$0.2 & 1314$\pm$390 & 107.5$\pm$12.1 \\
\hline
\hline
 & $i$ & $b_{1i}$ & $b_{2i}$ & $b_{5i}$ \\
\hline
$P$      & 1 & 3.8$\pm$0.14   & 7.45$\pm$0.42  & 2.67$\pm$0.25  \\
$f_2$    & 2 & 18.5$\pm$6.1  & 4.2$\pm$0.9 &  21.7$\pm$3.2 \\
$\omega$ & 3 & 0.28$\pm$0.05  &  0.12$\pm$0.04 & 0.0016$\pm$0.00003  \\
$\rho$   & 4 & 36$\pm$8  &  0.0007$\pm$0.0003 &   0.14$\pm$0.03  \\
$a_2$   & 5 & 0.57$\pm$0.06  &  1.66$\pm$0.23 &   1.42$\pm$0.13 \\
\hline
\end{tabular}
\end{center}
\end{table}
\renewcommand{\arraystretch}{1.0}

We use data at proton laboratory momenta within the range of
3 to 50~GeV/c (corresponding to invariant collision energies 
of $2.77{<}\sqrt{s}{<}10$ GeV) because in this region the
contributions from the $\omega$, $\rho$, $f_2$ and
$a_2$ Regge exchanges play an essential role. 
As is illustrated in Fig.~\ref{kinem2} at higher energies 
the contributions from those Regge trajectories become almost 
negligible and $pp$ scattering is dominated by single Pomeron exchange. 
Indeed at higher collision energies other dynamical mechanisms become 
very important which, at the same time, are irrelevant in the energy 
range indicated above. 
For example, two classes of Pomeron trajectories are introduced 
in \cite{Donnachie01,Donnachie04,Cudell05}, namely a soft Pomeron
and a hard Pomeron. 
Other investigations of $pp$ scattering consider also additional 
contributions~\cite{Donnachie84,Donnachie86,Avila,Martynov} 
besides the single Pomeron Regge pole, in order to explain the 
data on differential cross sections at momenta above $\simeq$50 GeV/c
such as double-Pomeron exchange, three-gluon exchange, etc.
See also Ref.~\cite{Krisch}. 
Since we are predominantly interested in fixing the contributions
from the mentioned well-established traditional Reggeon contributions 
we do not extend our analysis to such very high energies. In the
energy region considered by us a single Pomeron Regge pole is 
sufficient. 

\begin{figure}[t]
\vspace*{-6mm}
\centerline{\hspace*{5mm}\psfig{file=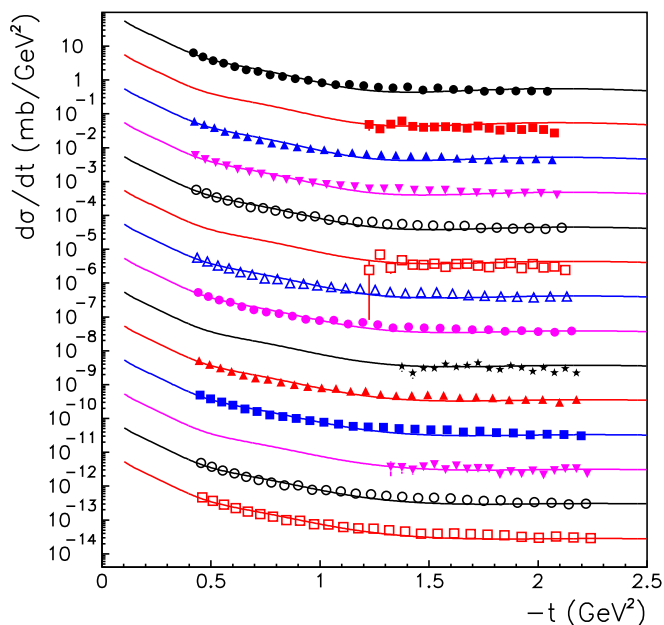,width=9.5cm}}
\vspace*{-5mm}
\caption{ \label{barp1} Differential cross section for $pp$ elastic 
scattering as a function of the four-momentum transfer squared $t$. 
Results are shown for different beam momenta (from top to bottom: 
3.012, 3.017, 
3.036, 3.062, 3.087, 3.095, 3.112, 3.137, 3.15, 3.162, 3.187, 3.205,
3.212 and 3.237 GeV/c). 
The solid lines are our model results. The data points and the lines are scaled
by factors of 10$^0$ (top) to 10$^{-13}$ (bottom), consecutively.
References to the data are given in Table~\ref{difcros}.
}
\end{figure}
\begin{figure}[h]
\vspace*{-6mm}
\centerline{\hspace*{5mm}\psfig{file=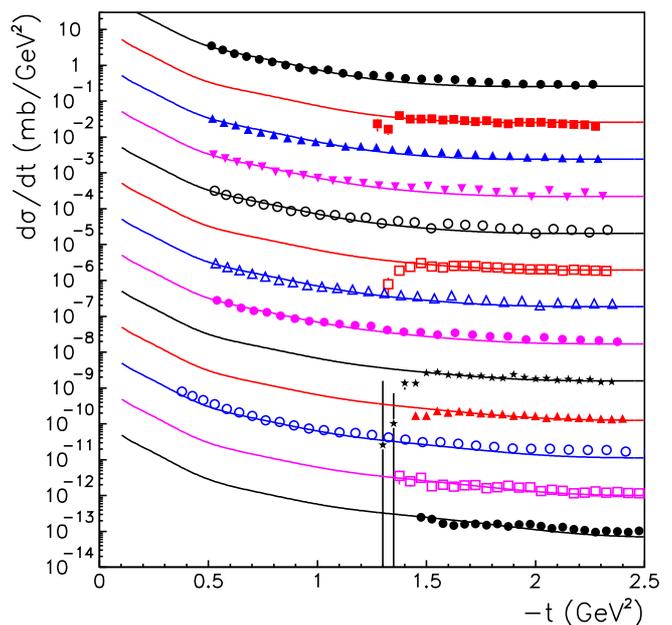,width=9.5cm}}
\vspace*{-5mm}
\caption{ \label{barp2} Differential cross section for $pp$ elastic 
scattering as a function of the four-momentum transfer squared $t$. 
Results are shown for different beam momenta (from top to bottom: 
3.262,
3.266, 3.287, 3.312, 3.337, 3.35, 3.362, 3.388, 3.41, 3.469, 3.499, 3.53,
3.621 GeV/c).
The solid lines are our model results. The data points and the lines are scaled
by factors of 10$^0$ (top) to 10$^{-11}$ (bottom), consecutively.
References to the data are given in Table~\ref{difcros}.
}
\end{figure}

It is not clear a priori down to which energies the Regge formalism is
applicable. Our previous systematic analyses of neutral and charged pion
photoproduction~\cite{Sibirtsev09a,Sibirtsev09,Sibirtsev03}, on pion
charge-exchange~\cite{Huang09} and pion-nucleon backward
scattering~\cite{Huang09a} demonstrate that Regge models allow to 
describe differential cross sections as well as single and double
polarization data fairly well down to $p{\simeq}3$ GeV/c. 
Thus, we decided to include in our fit also $pp$ 
scattering data down to $p{\simeq}3$ GeV/c.

Information about the data considered in the present analysis is
summarized in an Appendix. There we indicate the beam momenta, the maximal 
and minimal values of the four-momentum transfer squared, the 
group who performed the experiment and provide the reference of 
the publication. We include data in the range $|t|{\le}$2.5~GeV$^2$ which
is the range where most of the experiments were performed. 

In addition to the data listed in the Appendix we fit the total $pp$
cross section for momenta between 3 and 50~GeV/c and the ratio of 
the real to imaginary parts of the forward scattering amplitude. Applying
the optical theorem this allows us to fix the helicity non-flip amplitude 
at forward direction, {\it i. e.} at $t{=}0$. 
Moreover, in order to further constrain our amplitudes (but also to
test them) we include data on total cross sections for the 
${\bar p}p$, $pn$ and ${\bar p}p$ reactions too.
 
\section{Results}

\begin{figure}[t]
\vspace*{-6mm}
\centerline{\hspace*{5mm}\psfig{file=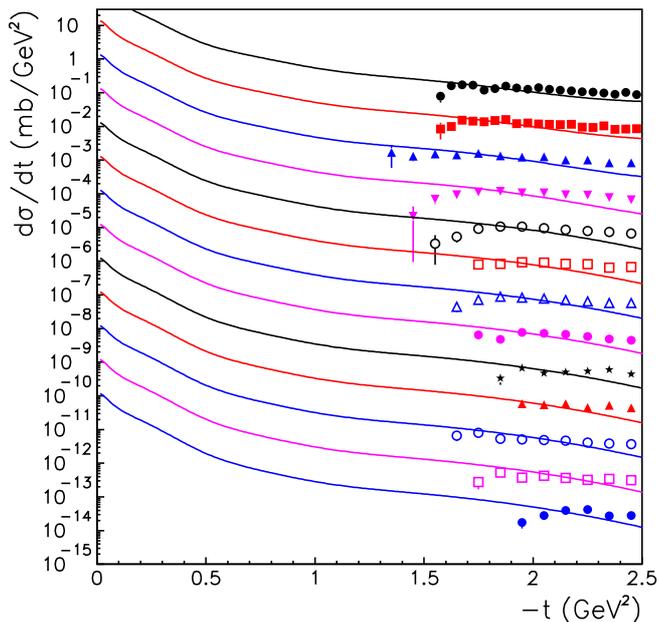,width=9.5cm}}
\vspace*{-5mm}
\caption{ \label{barp3} Differential cross section for $pp$ elastic 
scattering as a function of the four-momentum transfer squared $t$. 
Results are shown for different beam momenta (from top to bottom: 
 3.686, 3.75, 3.843, 3.942, 4.013, 4.082, 4.151, 4.258, 4.334, 4.409,
4.483, 4.559 and 4.681
GeV/c).
The solid lines are our model results. The data points and the lines are scaled
by factors of 10$^0$ (top) to 10$^{-12}$ (bottom), consecutively.
References to the data are given in Table~\ref{difcros}.
}
\end{figure}

\subsection{Differential cross sections}

\begin{figure}[b]
\vspace*{-6mm}
\centerline{\hspace*{5mm}\psfig{file=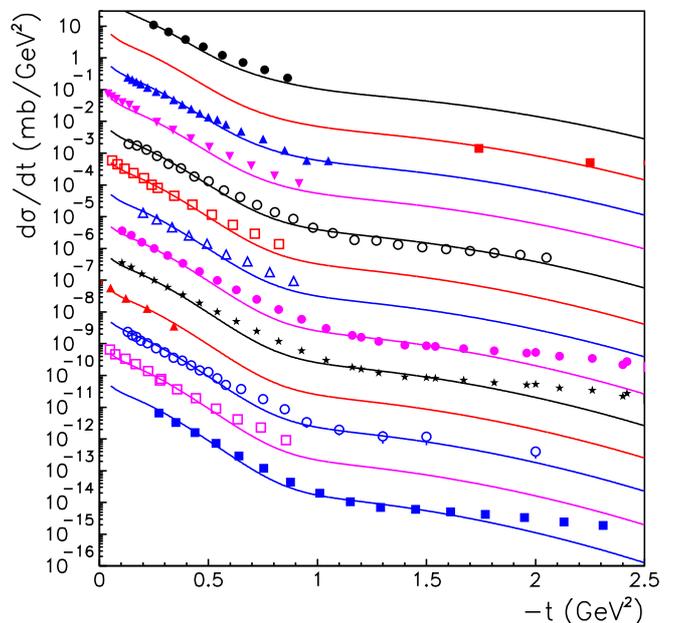,width=9.5cm}}
\vspace*{-5mm}
\caption{ \label{barp3a} Differential cross section for $pp$ elastic 
scattering as a function of the four-momentum transfer squared $t$. 
Results are shown for different beam momenta (from top to bottom: 
6.8, 8.0, 8.5, 8.8, 10.0 , 10.8, 10.94, 12.0, 12.0, 12.1, 12.4, 12.8, 14.2
GeV/c).
The solid lines are our model results. The data points and the lines are scaled
by factors of 10$^0$ (top) to 10$^{-13}$ (bottom), consecutively.
References to the data are given in Tables~\ref{difcros} and \ref{difcros1}.
}
\end{figure}

\begin{figure}[t]
\vspace*{-6mm}
\centerline{\hspace*{5mm}\psfig{file=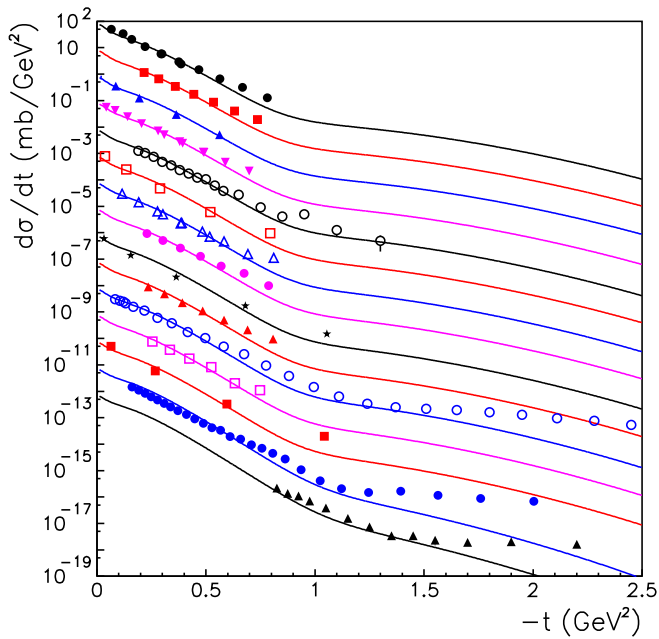,width=9.5cm}}
\vspace*{-5mm}
\caption{ \label{barp3b} Differential cross section for $pp$ elastic 
scattering as a function of the four-momentum transfer squared $t$. 
Results are shown for different beam momenta (from top to bottom: 
14.8, 14.93, 15.5, 16.7, 18.4, 18.6, 19.6, 19.84, 21.4, 21.88, 
24.0, 24.63, 26.2, 44.5, 50.0 GeV/c).
The solid lines are our model results. The data points and the lines are
scaled
by factors of 10$^0$ (top) to 10$^{-13}$ (bottom), consecutively.
References to the data are given in Tables~\ref{difcros} and \ref{difcros1}.
}
\end{figure}

\begin{figure}[t]
\vspace*{-6mm}
\centerline{\hspace*{5mm}\psfig{file=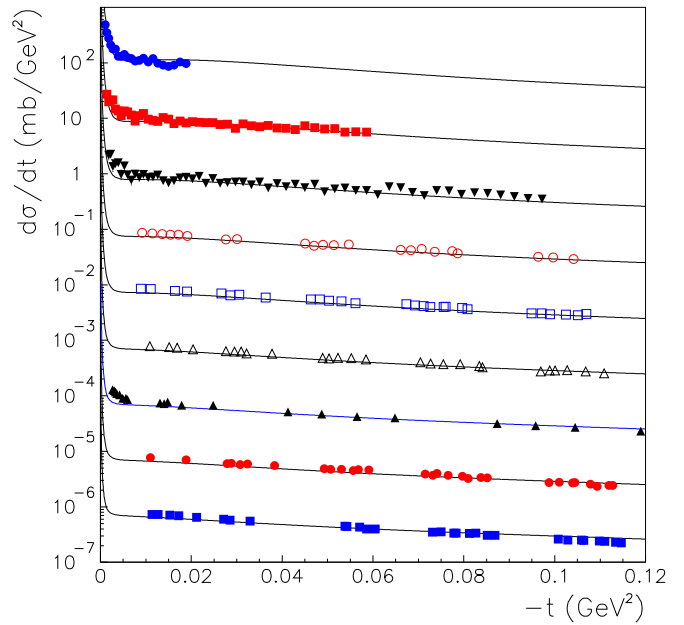,width=9.5cm}}
\vspace*{-5mm}
\caption{ \label{barp3c} Differential cross section for $pp$ elastic 
scattering as a function of the four-momentum transfer squared $t$. 
Results are shown for different beam momenta (from top to bottom: 
4.2, 7.0, 10.0, 13.16, 15.52, 24.56, 30.0, 30.45, and 45.17 GeV/c).
The solid lines are our model results. The data points and the lines are
scaled
by factors of 10$^0$ (top) to 10$^{-2}$ (bottom), consecutively.
References to the data are given in Tables~\ref{difcros} and
\ref{difcros1}.
}
\end{figure}

In Figs.~\ref{barp1}-\ref{barp3c} we show data on $pp$ scattering
differential cross sections at various beam momenta as indicated in the
figures. References to the data are given in 
Tables~\ref{difcros} and \ref{difcros1}. 
Most of the experimental results in the 
indicated kinematic region are available from the Zero Gradient Synchrotron
(ZGS) at Argonne National Laboratory and from the EDDA Collaboration at COSY
in J\"ulich. 
The results of our Regge model presented in Figs.~\ref{barp1}-\ref{barp3b}
are based on the central values of the parameters given in Table~\ref{param}. 
We do not display here the variations induced by the uncertainties in those 
parameters which are in the order of 20~\%, cf. Table~\ref{param}, in 
order to avoid too busy figures\footnote{
Note that more digits are needed for the parameter $c_2$ for the 
$\omega$ and $\rho$ exchange trajectories if one wants to reproduce
our results as given in Figs.~\ref{barp1}-\ref{barp3b} accurately, 
namely $c_{2\, \omega} = 376.117$ and $c_{2\, \rho} = -216.879$.}.

Obviously our Regge model yields a good overall reproduction of the 
differential cross sections over a wide energy range. In particular, 
it is possible to reproduce the experimental information for 
beam momenta from 3 to 50~GeV/c with the same set of parameters.
The obtained $\chi^2$ is with 1.69 per data point fairly good. Still
there is, in average, a discrepancy of about 20\% between 
our model results and some of the data, especially at the lower energies
considered. It is partly due to the known disagreement between some 
differential-cross-section data caused by differences in the absolute cross 
section normalization of various experiments, discussed in 
Ref.~\cite{Albers04}. As mentioned in this reference, the ZGS ANL
results~\cite{Jenkins78,Jenkins80} are lower by about 20\% with respect to
the most recent COSY-EDDA data~\cite{Albers04} and they also clearly disagree 
with previous measurements from ZGS ANL~\cite{Kammerud71}. 
In this context let us also mention that close to the maximal value of the 
squared four-momentum transfer $t$ accessible at the ZGS ANL experiment 
some of the differential cross section data deviate significantly 
from our calculations and seem to indicate an unexpected $t$-dependence, 
cf. Figs.~\ref{barp2}-\ref{barp3}.

The differential cross sections do not show any minima or other structures
that correspond to the zeros of the amplitudes due to exchanges
of individual trajectories as they are illustrated in Fig.~\ref{kinem2}. 
However, at the higher energies considered one can see the onset of
a shoulder around $t{\approx}{-}$1.4~GeV$^2$ 
(Figs.~\ref{barp3a}-\ref{barp3b}) which for very high energies evolves
into a dip structure, cf. Refs.~\cite{Donnachie84,Donnachie86,Avila,Martynov}.
Note that the more pronounced shoulder seen in the measurement at 
$p{=}44.5$~GeV/c \cite{Bruneton77}, seems to be in conflict with the bulk 
of the data as pointed out by E.~Martynov \cite{Martynov}.

As is clear from the data presented in 
Figs.~\ref{barp1}-\ref{barp2}, for momenta between 3 and 3.4~GeV/c 
there are almost no cross-section data available at $|t|{<}$0.5 GeV$^2$. 
Furthermore, in the momentum range 3.5${\le}p{<}$4.2 GeV/c there is 
basically no experimental information at $|t|{<}$1.5 GeV$^2$. 
Therefore, our amplitude in forward direction is fixed practically 
only by the data available at high momenta. Some data at higher
momenta and at very-forward angles are displayed in Fig.~\ref{barp3c}. 
Obviously, they are all very nicely reproduced by our Regge model. 
Note that here the contribution of the Coulomb amplitude \cite{Holzenkamp}, 
properly symmetrized, is included in the calculation. 

We expect that additional data at small $|t|$ will put more constraints 
on our solution and, therefore, would allow to determine better the reaction 
amplitude.
Hence, the upcoming data on differential cross sections from ANKE at COSY 
could play a decisive role for improving the analysis and for the extraction 
of high quality $pp$ scattering amplitudes.  As will be discussed in the next 
section, also the amount of polarization data for low ${-}t$ is rather 
limited. 
 
\subsection{Analyzing powers}
\begin{figure}[t]
\vspace*{-6mm}
\centerline{\hspace*{5mm}\psfig{file=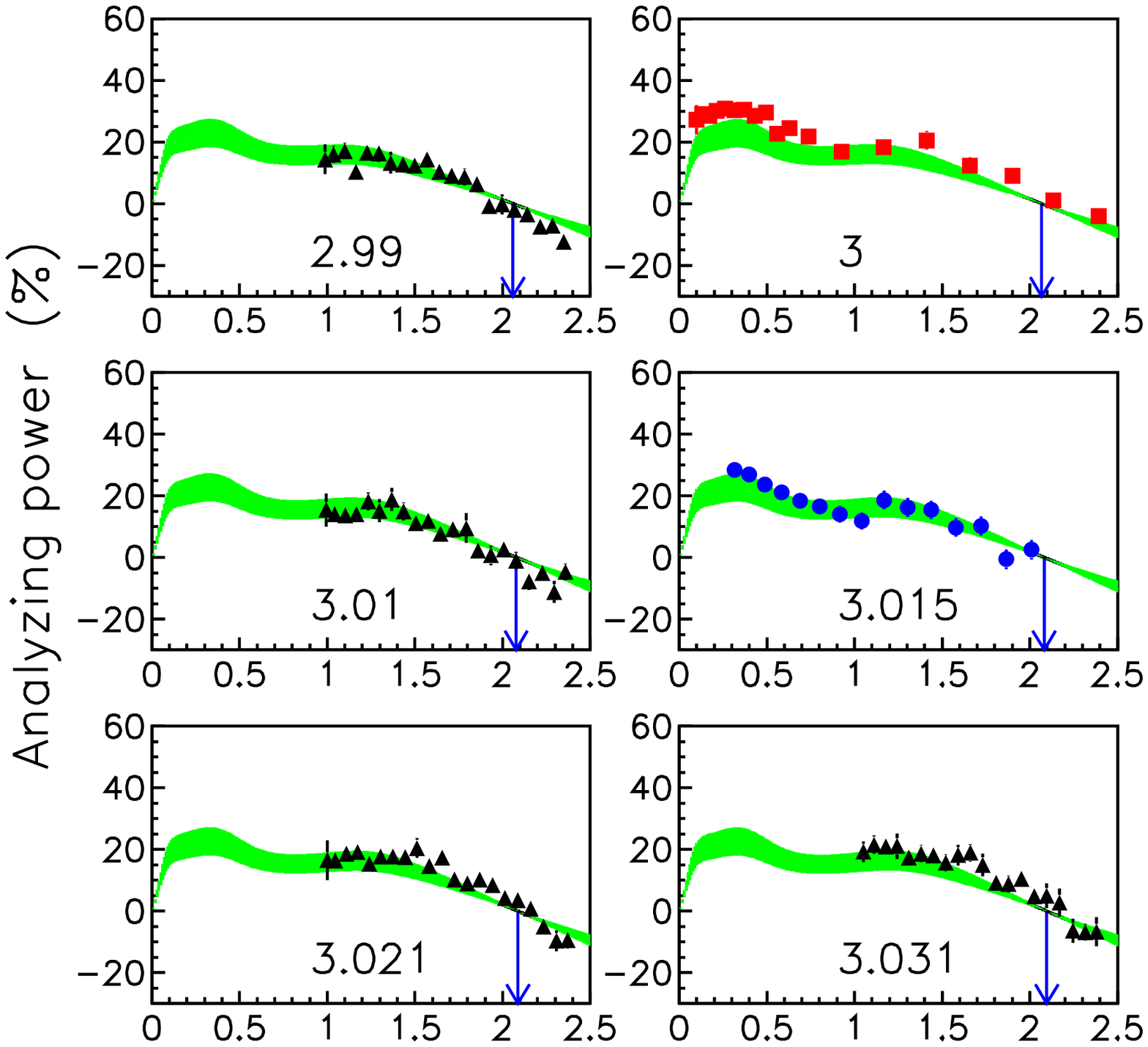,width=9.8cm}}
\vspace*{-17mm}
\centerline{\hspace*{5mm}\psfig{file=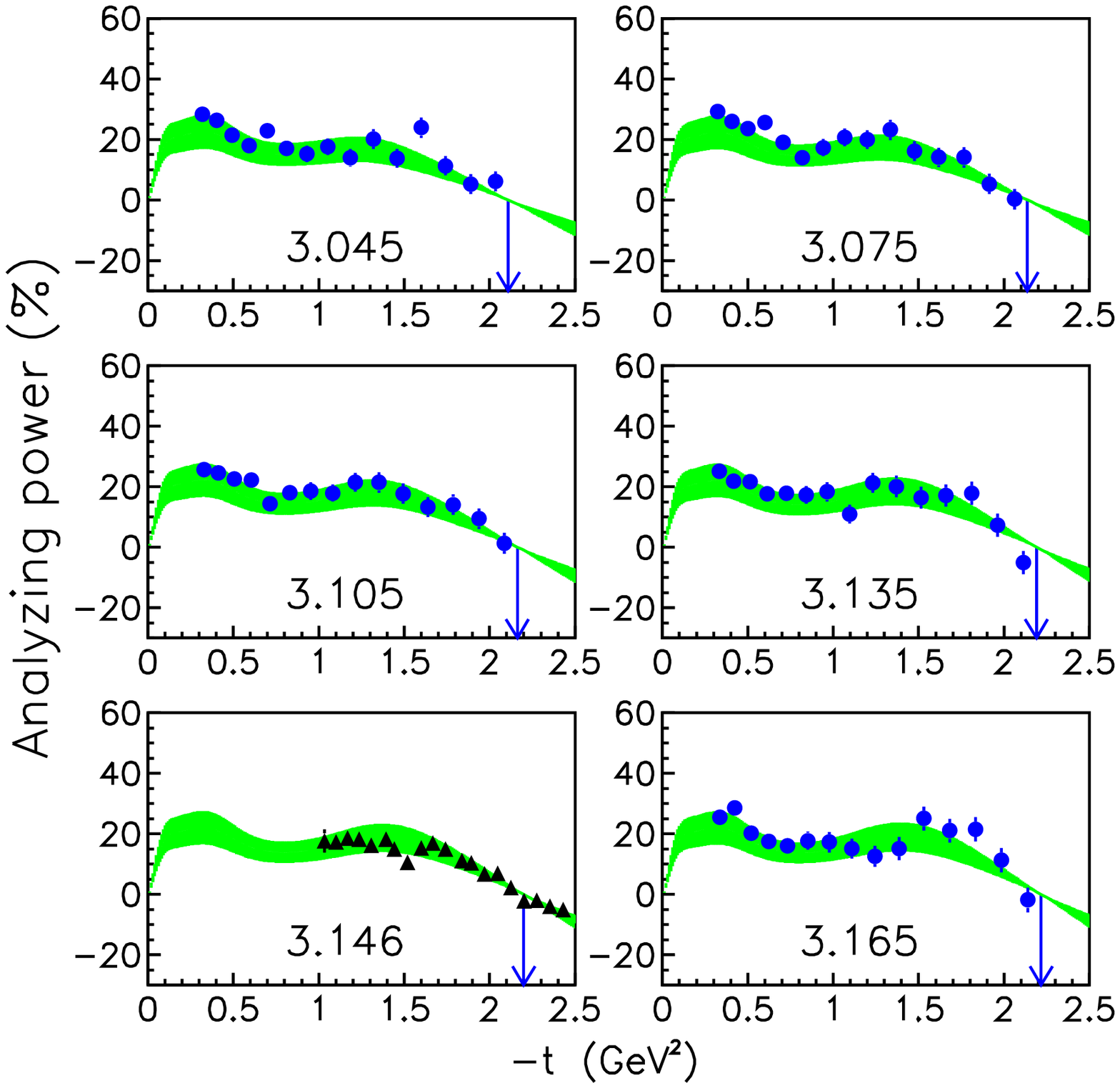,width=9.8cm}}
\vspace*{-5mm}
\caption{ \label{barp5} Analyzing power for $pp$ elastic scattering
as a function of the four-momentum transfer squared $t$ for different 
beam momenta (in GeV/c) as indicated in the figure. 
The triangles are data from SATURNE~\cite{Allgover99,Allgover99a},  
the squares are from ZGS ANL~\cite{Miller77}, and 
the circles are from COSY-EDDA~\cite{Altmeier00,Altmeier05}. 
The shaded bands illustrate the uncertainty of our calculation. 
The arrows indicate the squared four-momenta corresponding to the
scattering angle $\theta_{c.m.}{=}90^o$.
}
\end{figure}

\begin{figure}[t]
\vspace*{-6mm}
\centerline{\hspace*{5mm}\psfig{file=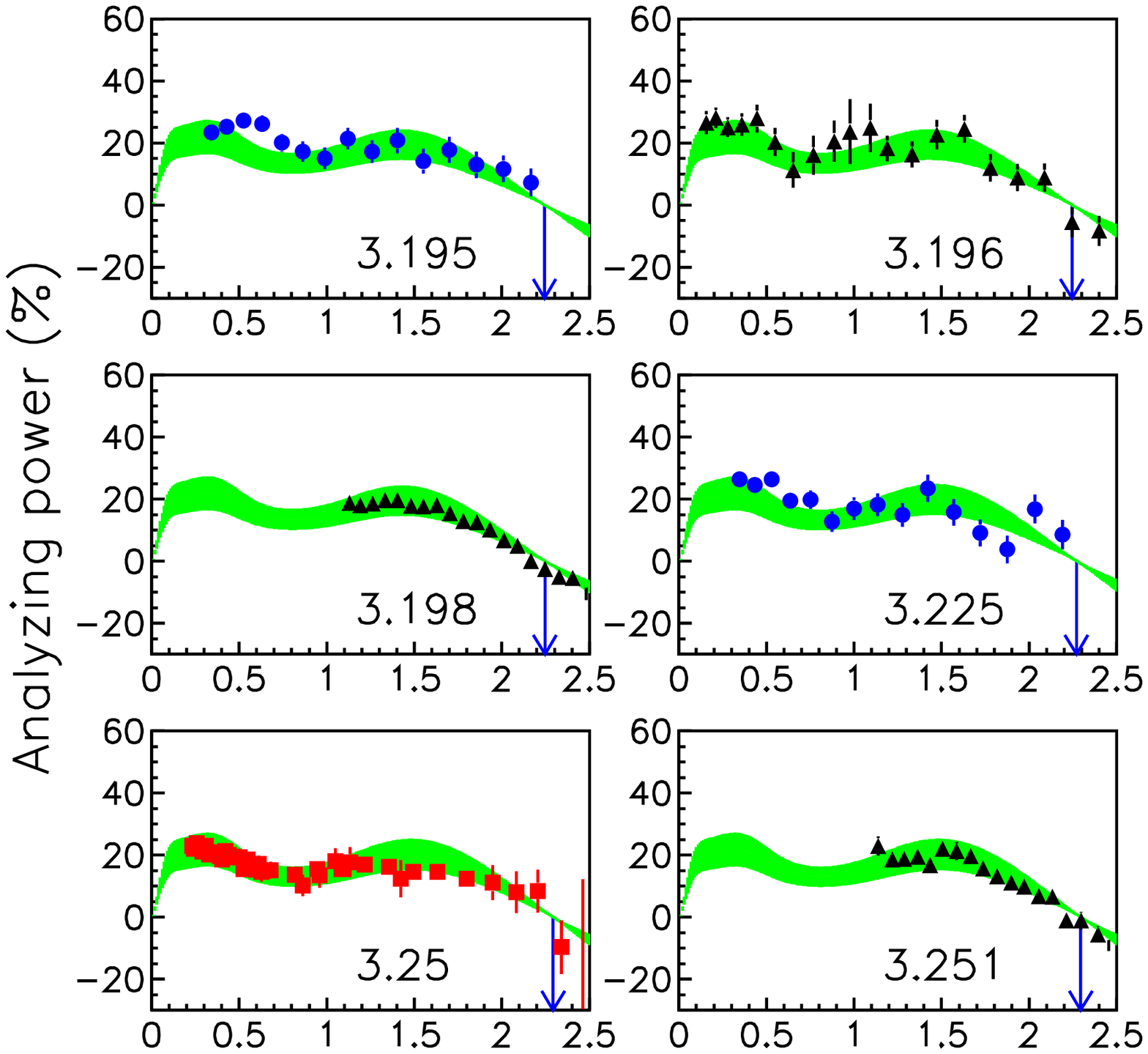,width=9.8cm}}
\vspace*{-17mm}
\centerline{\hspace*{5mm}\psfig{file=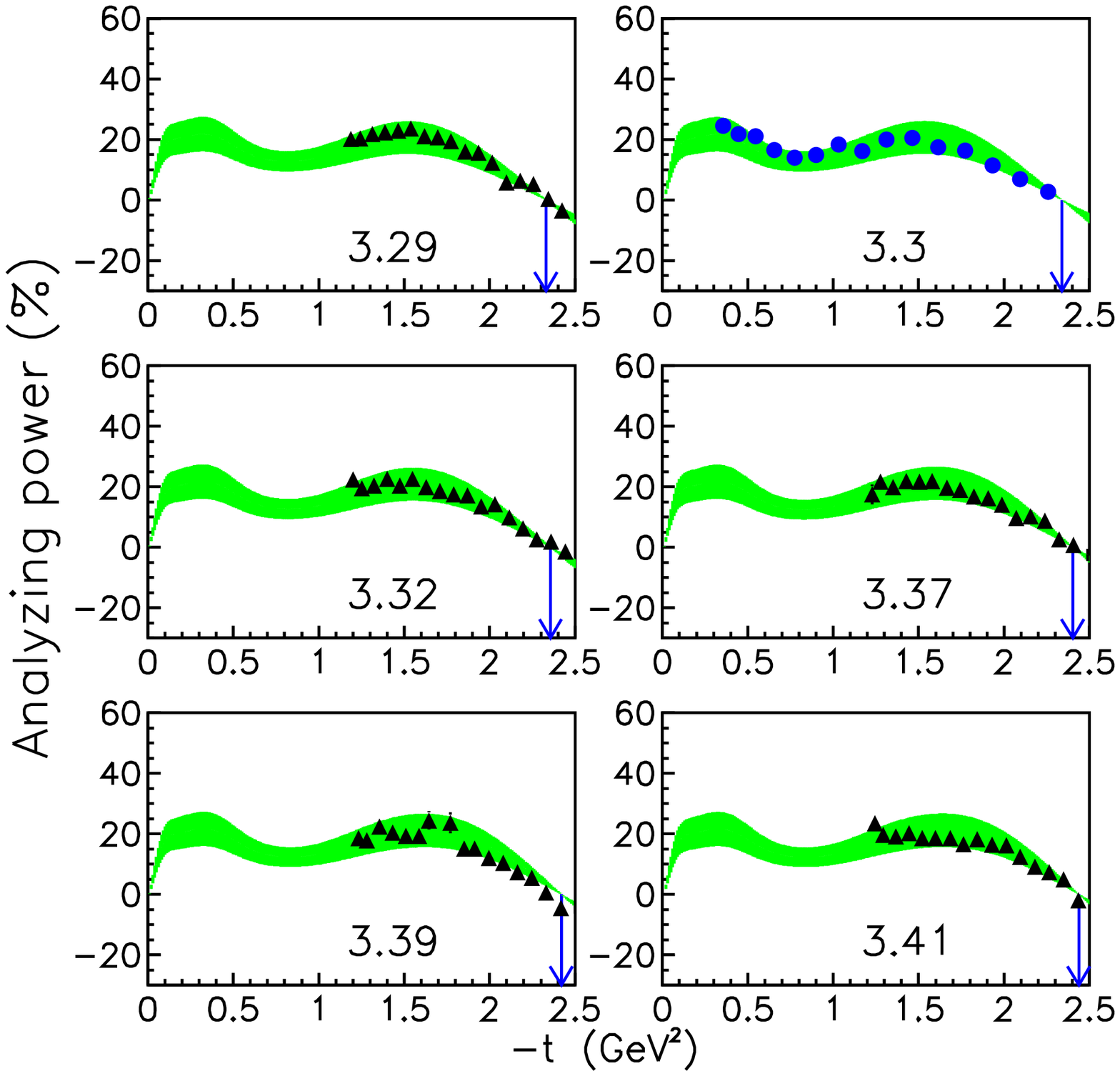,width=9.8cm}}
\vspace*{-5mm}
\caption{ \label{barp7} Analyzing power for $pp$ elastic scattering
as a function of the four-momentum transfer squared $t$ for different
beam momenta (in GeV/c) as indicated in the figure.
The triangles are data from SATURNE~\cite{Allgover99a,Perrot87},
the squares are from ZGS ANL~\cite{Parry73}, and
the circles are from COSY-EDDA~\cite{Altmeier00,Altmeier05}.
The shaded bands illustrate the uncertainty of our calculation.
The arrows indicate the squared four-momenta corresponding to the
scattering angle $\theta_{c.m.}{=}90^o$.
}
\end{figure}

In the following we use the notation given in Table~2 of 
Bystricky et al.~\cite{Bystricky78} for the polarization observables. 
Furthermore, we assume that $A{=}P$ which follows from time reversal
invariance and we do not specify explicitely which of those two 
observables was determined in the actual experiment. 
In the discussion below we refer to the proton polarization data 
always as analyzing powers.

\begin{figure}[t]
\vspace*{-6mm}
\centerline{\hspace*{5mm}\psfig{file=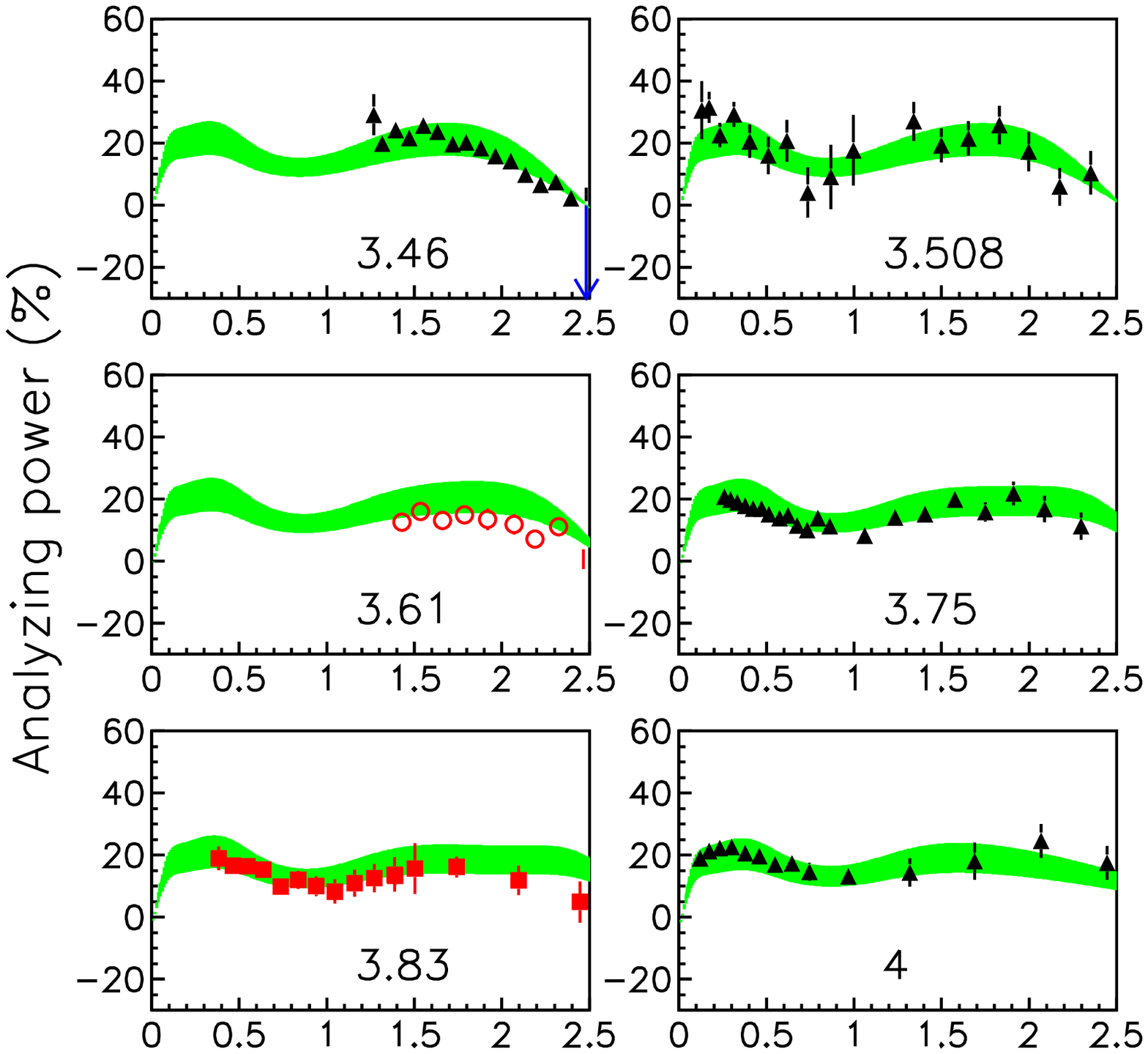,width=9.8cm}}
\vspace*{-17mm}
\centerline{\hspace*{5mm}\psfig{file=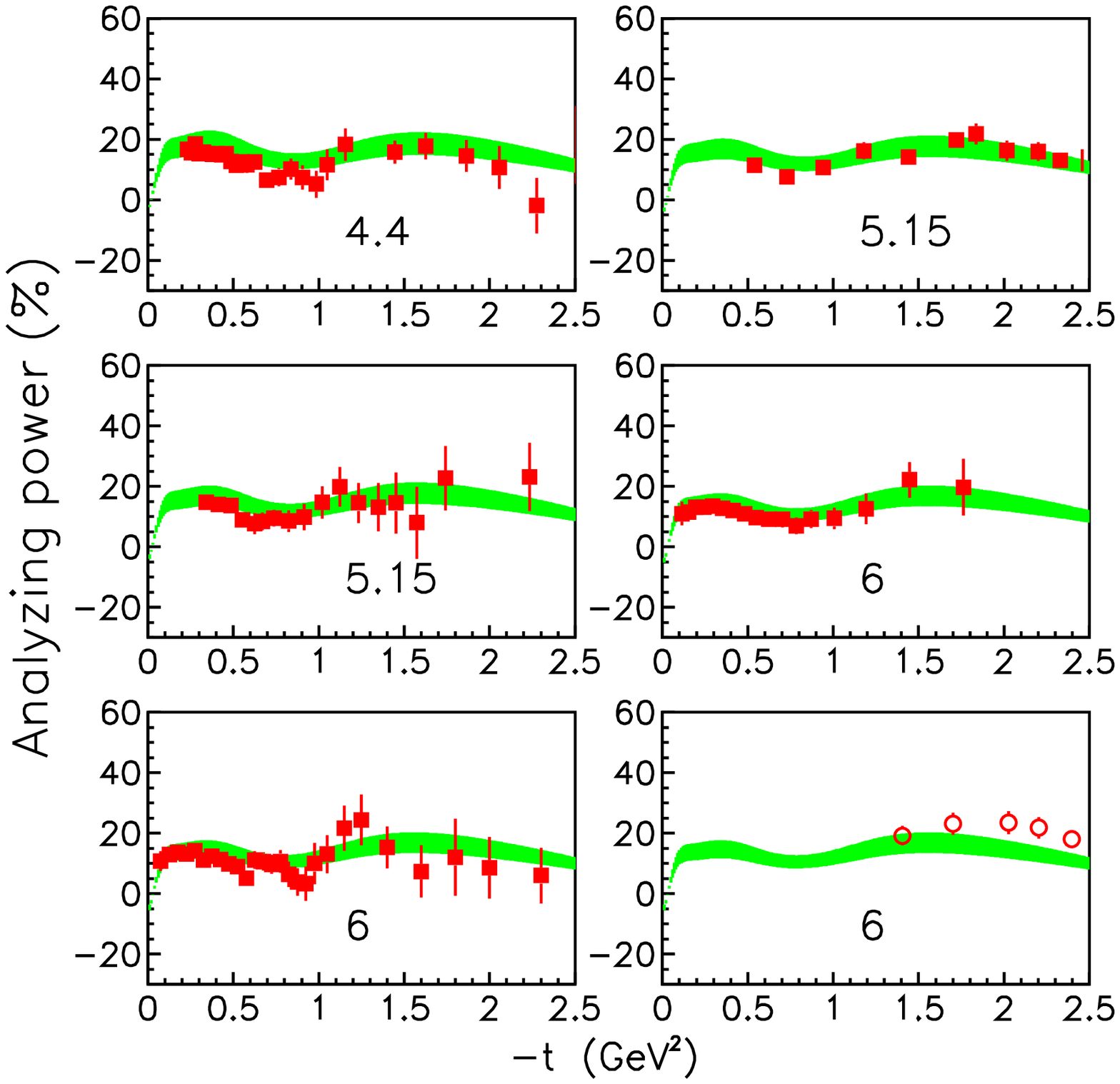,width=9.8cm}}
\vspace*{-5mm}
\caption{ \label{barp9} Analyzing power for $pp$ elastic scattering
as a function of the four-momentum transfer squared $t$ for different
beam momenta (in GeV/c) as indicated in the figure.
The triangles are data from SATURNE~\cite{Allgover99a,Perrot87,Deregel76},
the squares are from ZGS ANL~\cite{Miller77,Parry73,Abshire74}, and
the open circles are from CERN PS~\cite{Borghini70}. 
The shaded bands illustrate the uncertainty of our calculation.
The arrows indicate the squared four-momenta corresponding to the
scattering angle $\theta_{c.m.}{=}90^o$.
}
\end{figure}

\begin{figure}[t]
\vspace*{-6mm}
\centerline{\hspace*{5mm}\psfig{file=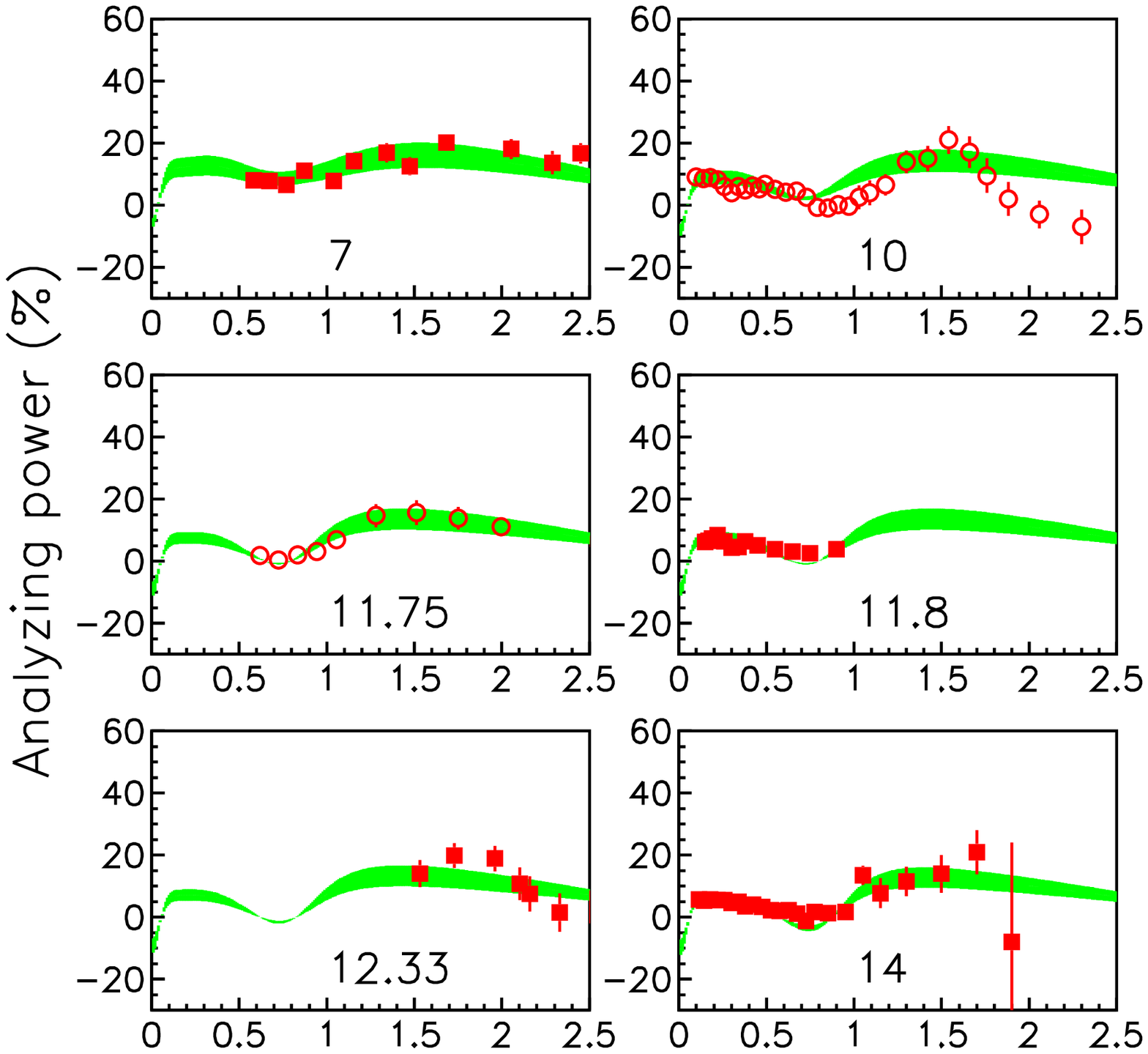,width=9.8cm}}
\vspace*{-17mm}
\centerline{\hspace*{5mm}\psfig{file=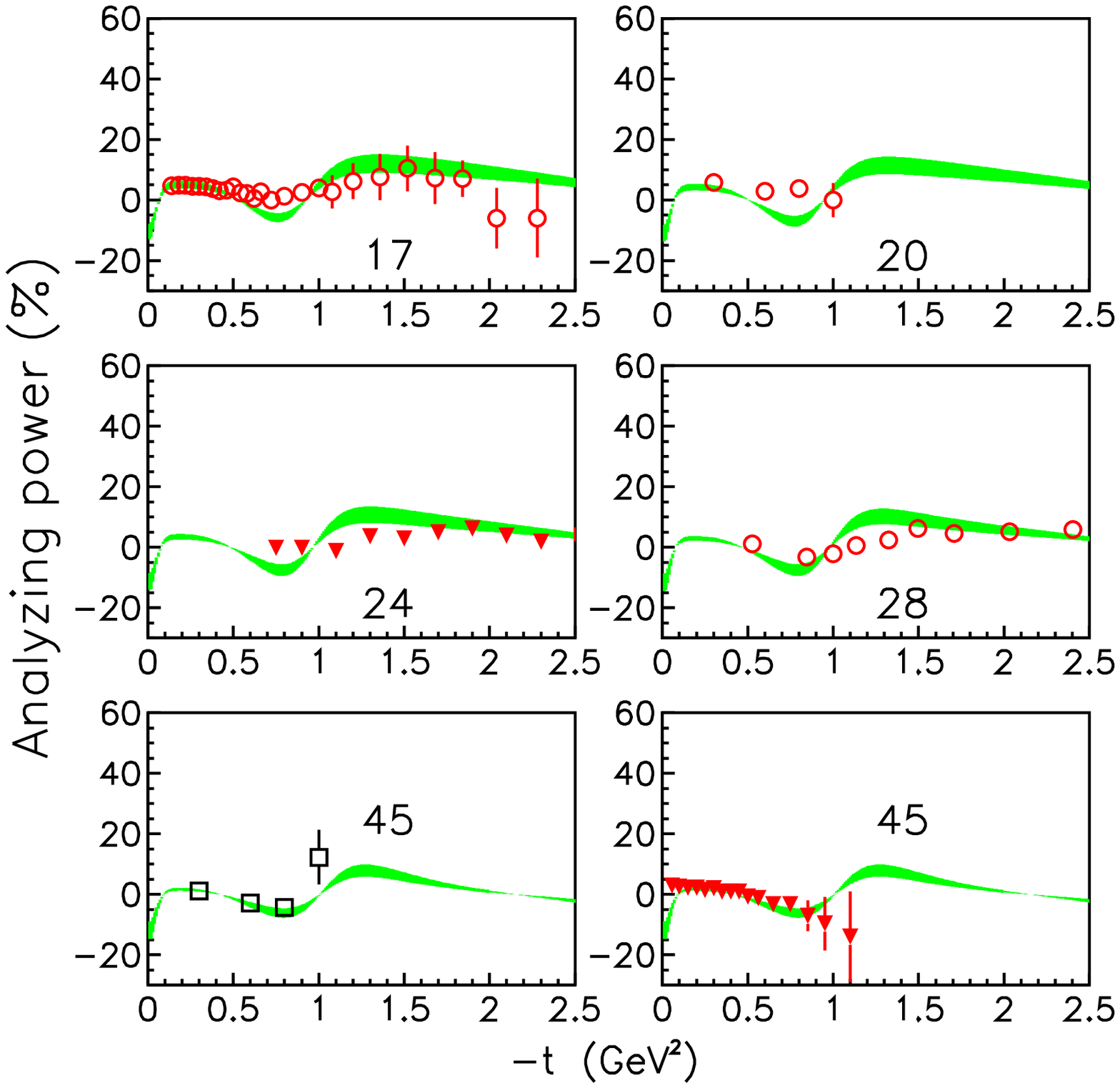,width=9.8cm}}
\vspace*{-5mm}
\caption{ \label{barp11} Analyzing power for $pp$ elastic scattering
as a function of the four-momentum transfer squared $t$ for different
beam momenta (in GeV/c) as indicated in the figure.
The inverse triangles are data from FNAL~\cite{Corcoran80},
the squares are from ZGS ANL~\cite{Abshire74,Linn82,Abe76,Kramer71}, and
the open circles are from CERN PS~\cite{Borghini71,Antille81}. 
The shaded bands illustrate the uncertainty of our calculation.
The arrows indicate the squared four-momenta corresponding to the
scattering angle $\theta_{c.m.}{=}90^o$.
}
\end{figure}

During the last years an extensive program on measuring analyzing
powers for $pp$ elastic scattering was conducted at
COSY-EDDA and at SATURNE. At the COSY facility data were
taken~\cite{Altmeier00,Altmeier05} at beam momenta from 1 to
3.3~GeV/c and for scattering angles from 30$^\circ$ to 90$^\circ$.
At SATURNE the measurements were
done~\cite{Allgover99,Allgover99a}
at beam momenta from 2.57 to 3.61~GeV/c and for proton scattering 
angles typically from 60$^\circ$ to 100$^\circ$. 
In addition, within this momentum range there are analyzing powers available 
from ZGS ANL~\cite{Jenkins78,Jenkins80}.  At higher momenta data are available 
from ZGS ANL, CERN PS, FNAL, and AGS BNL. References to the data are collected
in Table~\ref{apower}.

Figs.~\ref{barp5}-\ref{barp11} display $pp$ 
analyzing powers as a function of the four-momentum transfer squared for
different beam momenta. The symbols indicate the experimental results from
different measurements as explained in the figure captions. The
data are shown only with statistical errors. The shaded
bands illustrate the uncertainties of our calculation. 
Those bands are obtained by varying the values of the model parameters 
individually within the one standard deviation given in Table \ref{param}.
The arrows
indicate those values of $t$ which correspond to the $pp$ scattering angle 
of $\theta_{c.m.} =$ 90$^\circ$. There the analyzing powers equal to zero.

\begin{figure}[t]
\vspace*{-6mm}
\centerline{\hspace*{5mm}\psfig{file=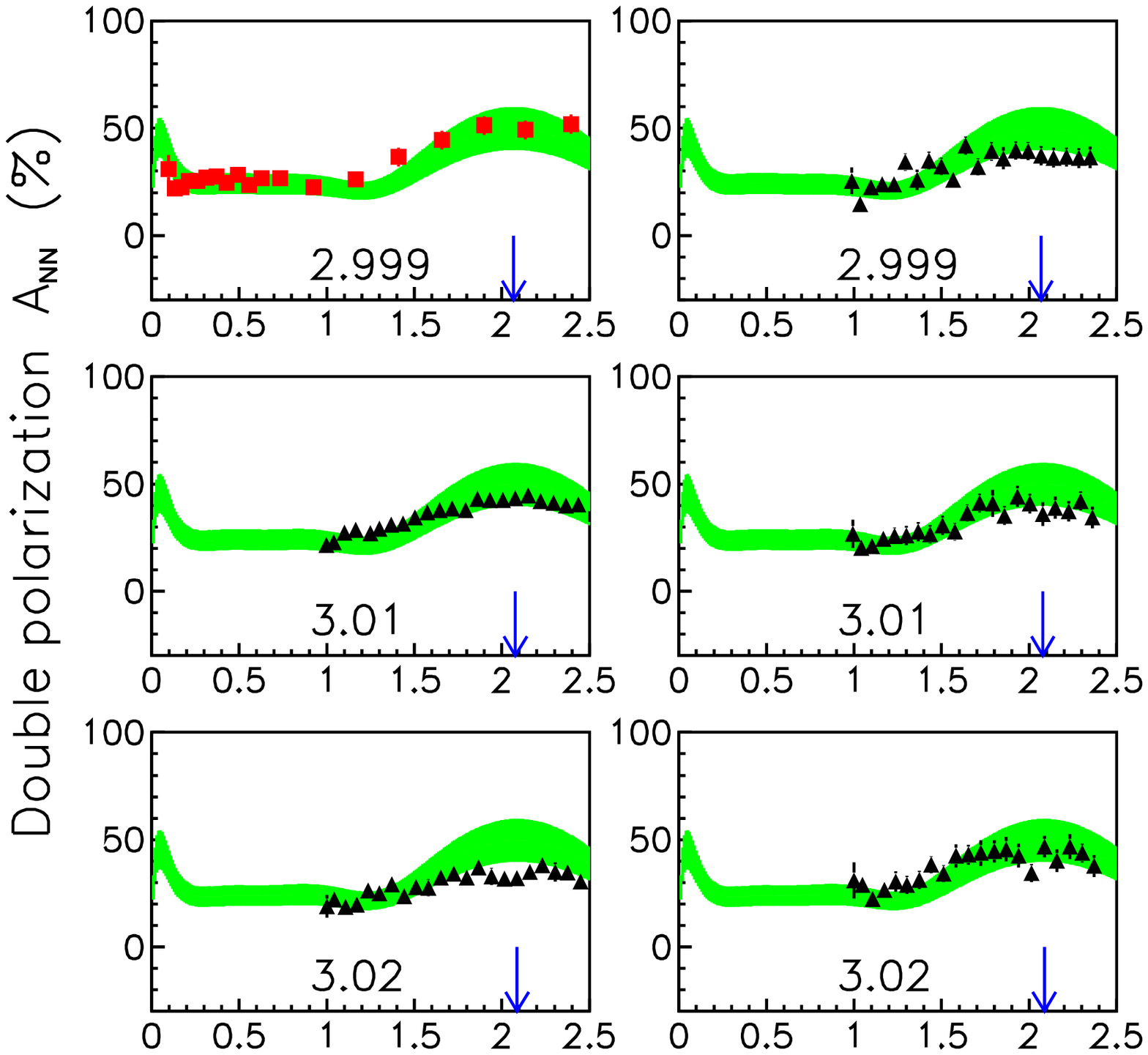,width=9.8cm}}
\vspace*{-17mm}
\centerline{\hspace*{5mm}\psfig{file=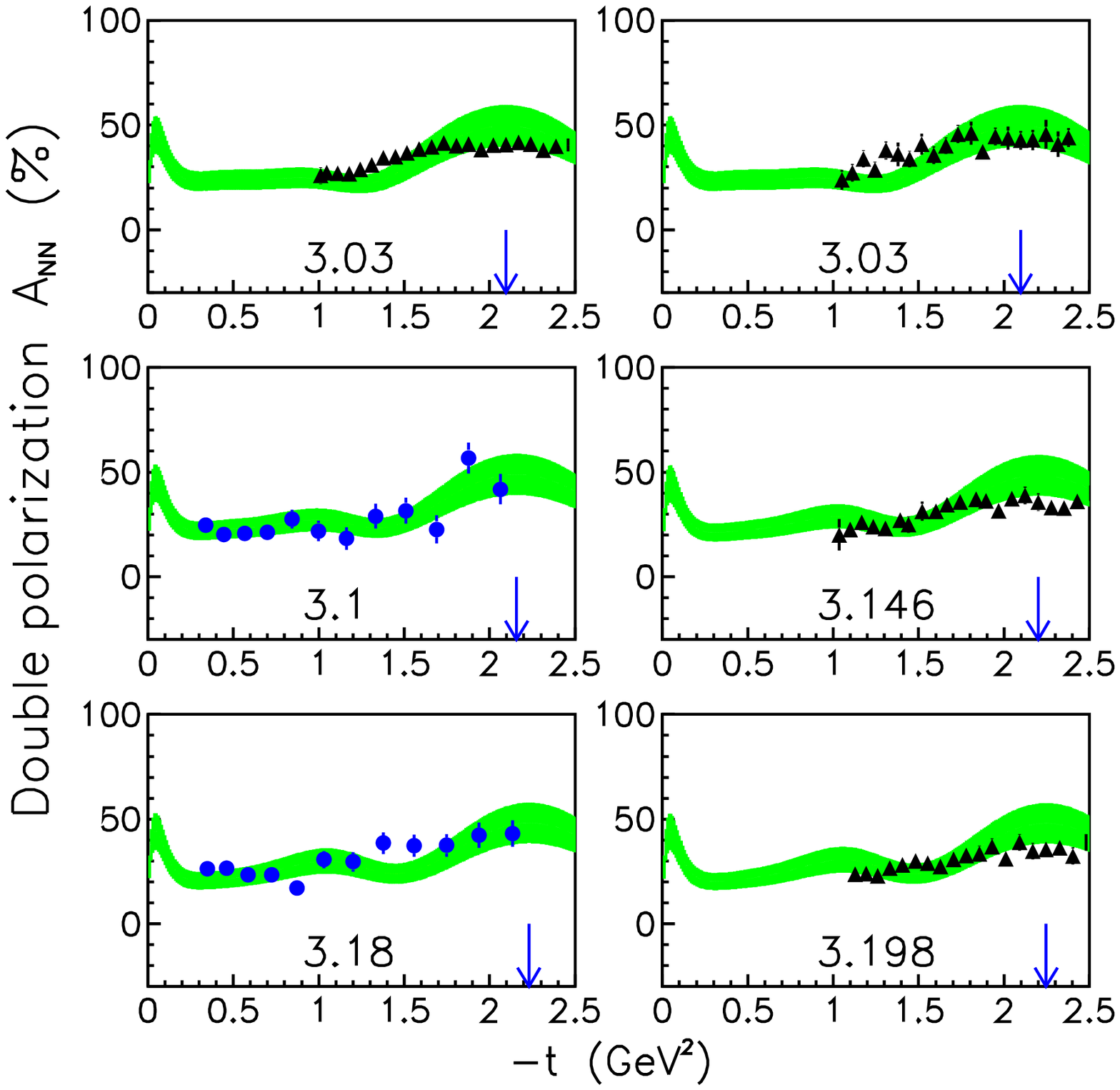,width=9.8cm}}
\vspace*{-5mm}
\caption{\label{ppyy2} Double polarization parameter $A_{NN}$
for $pp$ elastic scattering
as a function of the four-momentum transfer squared $t$ for different
beam momenta (in GeV/c) as indicated in the figure.
The triangles are data from SATURNE~\cite{Allgower00,Allgower01},  
the squares are from ZGS ANL~\cite{Miller77}, and 
the circles are from COSY-EDDA~\cite{Bauer05}. 
The shaded bands illustrate the uncertainty of our calculation.
The arrows indicate the squared four-momenta corresponding to the
scattering angle $\theta_{c.m.}{=}90^o$.
}
\end{figure}

Although the COSY-EDDA and SATURNE measurements provide many precise data these
experiments do not cover forward angles and, therefore, the behavior of 
the amplitudes for scattering angles below 30$^\circ$ is not well constrained 
experimentally. 

\begin{figure}[t]
\vspace*{-6mm}
\centerline{\hspace*{5mm}\psfig{file=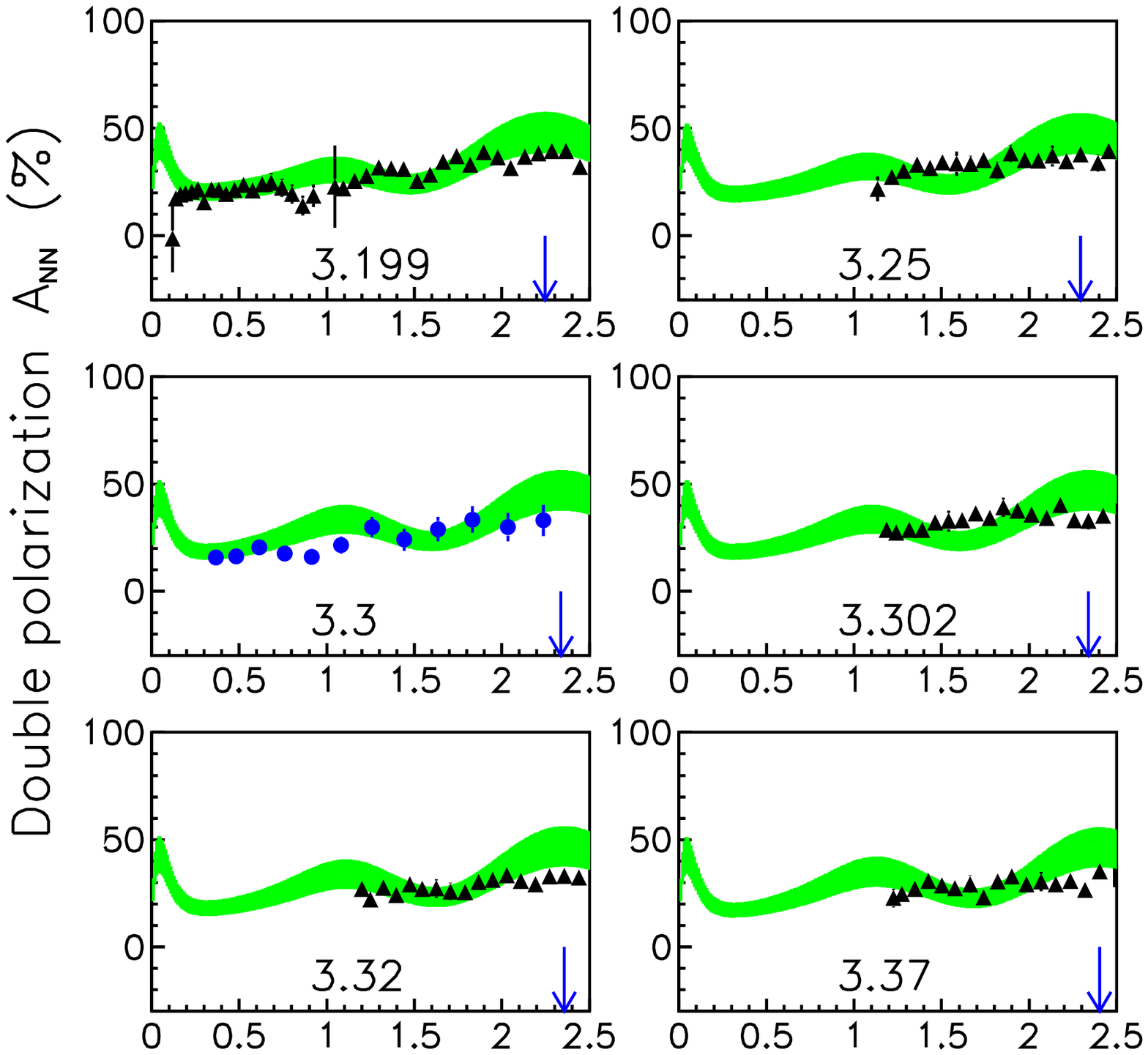,width=9.8cm}}
\vspace*{-17mm}
\centerline{\hspace*{5mm}\psfig{file=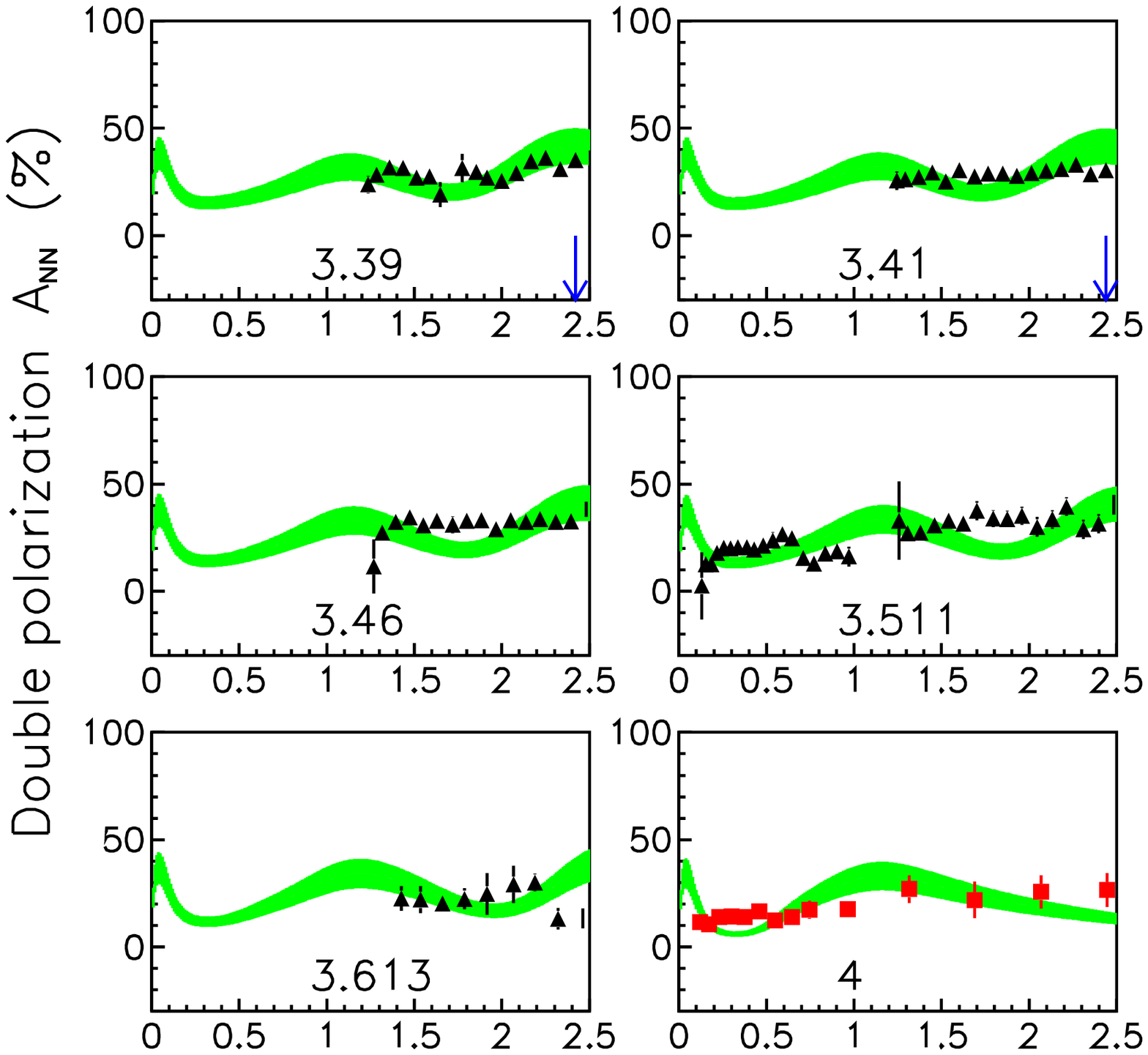,width=9.8cm}}
\vspace*{-5mm}
\caption{\label{ppyy4} Double polarization parameter $A_{NN}$
for $pp$ elastic scattering
as a function of the four-momentum transfer squared $t$ for different
beam momenta (in GeV/c) as indicated in the figure.
The triangles are data from SATURNE~\cite{Allgower00,Allgower01,Lehar87},  
the squares are from ZGS ANL~\cite{Miller77}, and 
the circles are from COSY-EDDA~\cite{Bauer05}. 
The shaded bands illustrate the uncertainty of our calculation.
The arrows indicate the squared four-momenta corresponding to the
scattering angle $\theta_{c.m.}{=}90^o$.
}
\end{figure}

The Regge calculation produces a minimum structure in the analyzing 
power around ${-}t{\simeq}0.5{\div}1$~GeV$^2$. The 
COSY-EDDA~\cite{Altmeier00,Altmeier05} 
and ZGS ANL~\cite{Jenkins78,Jenkins80} data available in that region
seem to support such a structure. But there are some beam momenta 
where the COSY data do not really exhibit a minimum. 

It is interesting that the analyzing powers do not vanish at high momenta.
Indeed, the data as well as our calculation exhibit some structure. 
The CERN
PS data~\cite{Borghini71} at the beam momentum of 10 GeV/c (cf. 
Fig.~\ref{barp11}) indicate a distinct maximum of the analyzing power around
${-}t{\simeq}1.6$ GeV$^2$. Our Regge model does not produce such a pronounced 
structure. Other data available at near-by beam momenta do not show such a 
pronounced structure either.

\subsection{Double polarization parameter $A_{NN}$}
The double polarization parameters discussed here
can be measured with a vertically polarized beam and a vertically polarized
target where the direction is defined with respect to the proton beam. 
Extensive measurements of $A_{NN}$ were done
at SATURNE~\cite{Allgower00,Allgower01,Lehar87} and at 
COSY-EDDA~\cite{Bauer05}. The double polarization observable $C_{NN}$, or spin
correlation parameter, was measured~\cite{Miller77} at ZGS ANL. Since 
$A_{NN}=C_{NN}$ due to time reversal invariance, in the following 
we use always the notation $A_{NN}$. References to the data considered in 
the present work are summarized in Table~\ref{ann}.

Most of the $A_{NN}$ data at beam momenta above 3~GeV/c come from
SATURNE. These data are shown by solid triangles in Figs.~\ref{ppyy2}
and \ref{ppyy4}. Note that at some beam momenta there are two data sets 
which were actually published in different papers and obviously 
obtained from different series of measurements, cf. Table~\ref{ann}. 
Unfortunately, the more recent SATURNE measurement~\cite{Allgower00,Allgower01} 
does not cover the region of forward angles. In general, 
the SATURNE data do not indicate any structures. 
Our Regge model predicts some structure in the $t$-dependence of
the double polarization parameter
$A_{NN}$ at $|t|{\le}1\,$GeV$^2$, specifically a clear minimum around
$|t|{=}0.5\,$GeV$^2$. The COSY-EDDA data~\cite{Bauer05} at 3.1 GeV/c seem
to support such a minimum, though this is not the case for their results at 
two other momenta. 

\subsection{Total cross sections}
\begin{figure}[t]
\vspace*{-4mm}
\centerline{\hspace*{5mm}\psfig{file=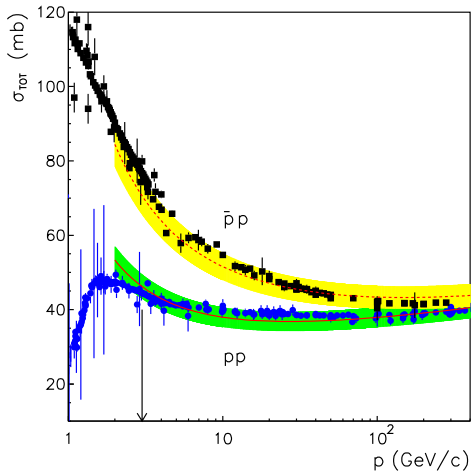,width=7cm}}
\vspace*{-3mm}
\caption{\label{pto1a}Total cross sections for the $pp$ and ${\bar p}p$
reactions as a function of the beam momentum. The data are taken
from the PDG~\cite{PDG}. The solid and dashed lines show our result
obtained via the optical theorem for $pp$ and ${\bar p}p$ scattering,
respectively. The shaded band illustrates the uncertainty of our model
calculation. The arrow indicates the beam momentum of 3 GeV/c. Only data
above that value were included in our fit.}
\end{figure}

The relation between the total $pp$ cross section and the
invariant helicity non-flip amplitudes is given by the optical theorem 
~\cite{Byckling73},
\begin{eqnarray}
\sigma_{tot}=\frac{{\rm Im} [\phi_1(t{=}0)+\phi_3(t{=}0)]}{2 \sqrt{s^2-4sm_N^2}}~. 
\end{eqnarray}
Data on the total $pp$ cross section, taken from the publication of the 
Particle Data Group~\cite{PDG}, are shown in Fig.~\ref{pto1a} as a 
function of the proton beam momentum. Note that in the global fit we included 
the data available at momenta from 3 to 50~GeV/c. 
However, as seen in Fig.~\ref{pto1a}, the total $pp$ cross section can be well 
described up to proton beam momenta of 400 GeV/c. In fact, the choice of the
intercept of the Pomeron contribution, cf. Eq.~(\ref{intercept}), implies that 
the $pp$ cross section predicted by our Regge model still remains in line with 
the data at much higher energies \cite{Donnachie92}. 
The arrow in Fig.~\ref{pto1a} indicates
the beam momentum of 3~GeV/c. As one can see, the total cross section is
reasonably well reproduced down to $p{\simeq}$2~GeV/c within the
indicated uncertainty.

\begin{figure}[h]
\vspace*{-4mm}
\centerline{\hspace*{5mm}\psfig{file=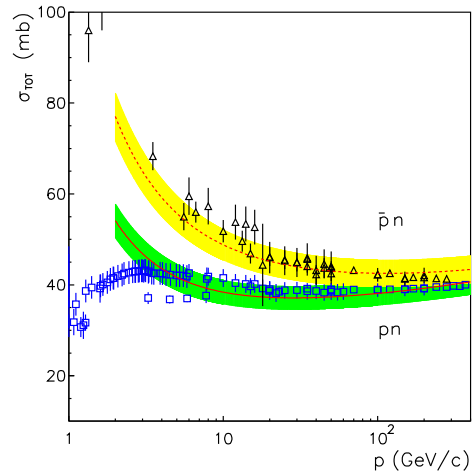,width=7cm}}
\vspace*{-3mm}
\caption{\label{pto1c} Total cross sections for the $pn$ and
${\bar p}n$ (c) as a function of the beam momentum shown by the solid and
dashed lines, respectively.
The data are taken from the PDG~\cite{PDG}.
}
\end{figure}

 \begin{figure}[b]
\vspace*{-4mm}
\centerline{\hspace*{5mm}\psfig{file=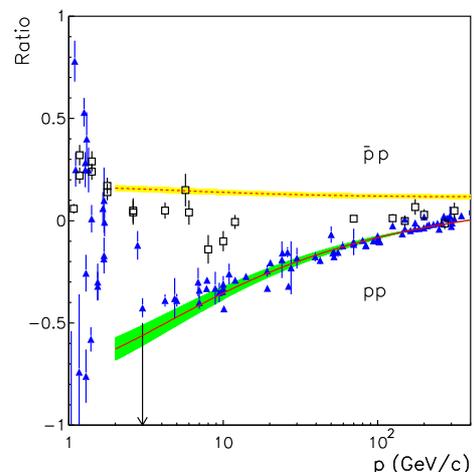,width=7cm}}
\vspace*{-3mm}
\caption{\label{pto1b} Ratio of the real-to-imaginary parts of the 
forward amplitudes for
$pp$ (triangles, solid line) and ${\bar p}p$ (squares, dashed line), 
respectively. The data are taken from the PDG~\cite{PDG}.
}
\end{figure}

Let us now present our results for the other baryon-baryon reactions 
involving nucleons and/or antinucleons. 
Taking into account the isospin structure and the $G$-parity relations
the contributions of the $\omega$, $\rho$, $f_2$, $a_2$ Regge 
exchanges and of the Pomeron to the 
$pp$, ${\bar p}p$, $pn$ and ${\bar p}p$ scattering amplitudes 
are given by (see, e.g. \cite{Pelaez06})
\begin{eqnarray}
 \phi(pp)&=&-\phi_\omega-\phi_\rho+\phi_{f_2}+\phi_{a_2}+\phi_P,
\nonumber \\
 \phi({\bar p}p)&=&\phantom{-} 
 \phi_\omega+\phi_\rho+\phi_{f_2}+\phi_{a_2}+\phi_P, 
\nonumber \\
 \phi(pn)&=&-\phi_\omega+\phi_\rho+\phi_{f_2}-\phi_{a_2}+\phi_P, 
\nonumber \\
 \phi({\bar p}n)&=&\phantom{-}
\phi_\omega-\phi_\rho+\phi_{f_2}-\phi_{a_2}+\phi_P.
\label{cc}
\end{eqnarray}
Corresponding results for the ${\bar p}p$, $pn$ and ${\bar p}p$ total cross 
sections can be found in Figs.~\ref{pto1a} and
\ref{pto1c}, together with pertinent data. 
Obviously, our Regge model yields a good overall description of those
three reaction channels too, over practically the whole considered momentum
range. 

Finally, Fig.~\ref{pto1b} shows the ratio of the real-to-imaginary parts of the
forward $pp$ and $\bar pp$ scattering amplitudes as a function of beam momentum. 
Again the data are taken from the PDG~\cite{PDG}. The $pp$ data are well described 
up to momenta of around 200 GeV/c. In case of $\bar pp$ our results seem to 
overestimate
the data somewhat, but one has to keep in mind that the experimental information 
is fairly poor over the whole considered momentum range. 

\section{Comparison with results from the GWU partial wave analysis}
It is interesting to compare our calculations with the results from
partial wave analyses. Among different PWA's the one from the George 
Washington University~\cite{Arndt97,Arndt00,Arndt07} extends up to 
proton beam momenta of about 3.8~GeV/c.

Since 
$pp$ elastic scattering is governed by five independent
amplitudes a determination of those amplitudes (modulo an overall
phase) would require 9 independent observables.
When projecting those amplitudes to partial waves one needs to
know them over the full angular range. However, the only observables 
at beam momenta above 3~GeV/c, where more or less full information 
on their angular dependence is available, 
are differential cross sections, analyzing powers (polarizations) and 
the double polarization parameter $A_{NN}=C_{NN}$. 
There are some data on other polarization observables, though they 
are available only for a few beam momenta above 3~GeV/c.
Therefore, obtaining a PWA solution at $p{\ge}$3~GeV/c that is unique
and represents the experimental results in an appropriate way is a
challenging task. Indeed the recent data from COSY-EDDA and SATURNE
were quite instrumental to improve the 
GWU PWA~\cite{Arndt97,Arndt00,Arndt07}, available at the 
SAID~\cite{Arndt83,Arndt94} webpage.

\begin{figure}[t]
\vspace*{-5mm}
\centerline{\hspace*{5mm}\psfig{file=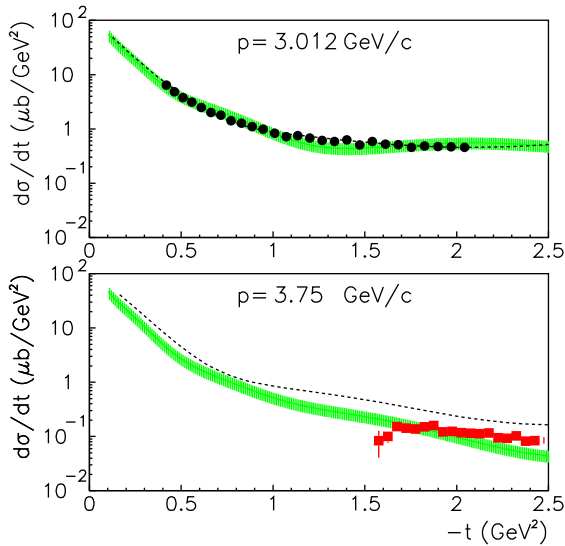,width=8cm}}
\vspace*{-4mm}
\caption{\label{compa1} Differential cross section for $pp$ 
scattering as a function of the four-momentum transfer squared 
for the beam momenta of 3.012 GeV/c and 3.75 GeV/c. 
The circles are 
data from COSY-EDDA~\cite{Albers04}, while the squares are the results
from ZGS ANL~\cite{Jenkins78,Jenkins80}. The shaded bands
illustrate the uncertainty of our calculation. The dashed lines 
represent the results from the GWU PWA~\cite{Arndt00,Arndt07}.}
\end{figure}

\begin{figure}[h]
\vspace*{-5mm}
\centerline{\hspace*{1mm}\psfig{file=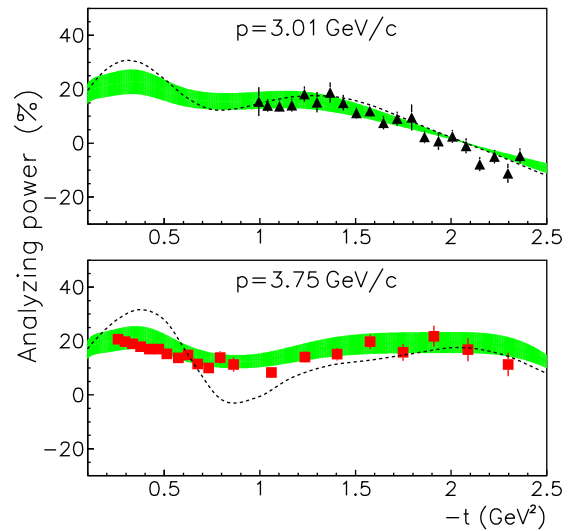,width=8cm}}
\vspace*{-4mm}
\caption{\label{compa2} Analyzing power for $pp$ 
scattering as a function of the four-momentum transfer squared for the
beam momenta of 3.01 GeV/c and 3.75 GeV/c. The triangles are 
data from SATURNE~\cite{Allgover99}, while the squares are the results
from ZGS ANL~\cite{Parry73}. The shaded bands
illustrate the uncertainty of our calculation. The dashed lines 
represents the results from the GWU PWA~\cite{Arndt00,Arndt07}.}
\end{figure}

Since the GWU PWA and our Regge analysis overlap within the momentum range
of 3 to 3.8~GeV/c we show here selected results close to these
two momenta for a comparison. For illustration we also include the 
experimental results.

\begin{figure}[t]
\vspace*{-5mm}
\centerline{\hspace*{1mm}\psfig{file=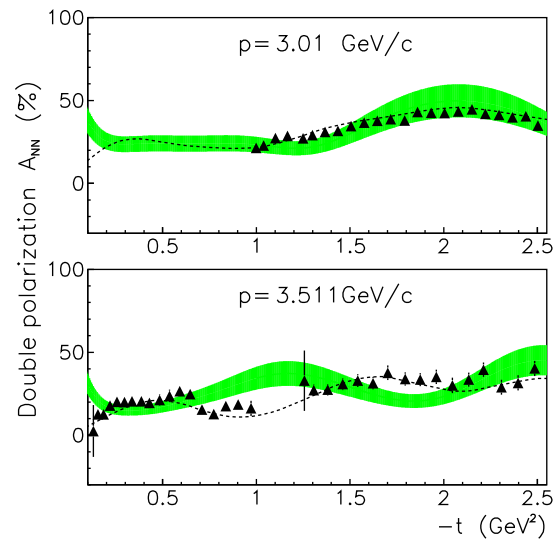,width=8cm}}
\vspace*{-4mm}
\caption{\label{compa3} Double polarization parameter $A_{NN}$ for
$pp$ scattering as a function of the four-momentum transfer squared 
for the beam momenta of 3.01 GeV/c and 3.511 GeV/c. The triangles are 
data from SATURNE~\cite{Allgower01,Lehar87}. The shaded bands
illustrate the uncertainty of our calculation. The dashed lines 
represents the results from the GWU PWA~\cite{Arndt00,Arndt07}.}
\end{figure}

The differential cross sections for $pp$ scattering at beam momenta of 
3.012 and 3.75~GeV/c are displayed in Fig.~\ref{compa1}. The data 
are from measurements at COSY-EDDA~\cite{Albers04} and at ZGS
ANL~\cite{Jenkins78,Jenkins80}. Again the shaded bands indicate the
uncertainty of our solution. The dashed lines are results based
on the GWU PWA~\cite{Arndt00,Arndt07}.
At the lower beam momentum our Regge model as well 
as the GWU PWA are in line with the experimental information
while at the higher momentum the GWU PWA overestimates the
data somewhat. 

Fig.~\ref{compa2} depicts the analyzing powers for $pp$ 
scattering at beam momenta of 3.01 and 3.75~GeV/c. The data are
from measurements at SATURNE~\cite{Allgover99} and at ZGS
ANL~\cite{Parry73}. 
At the beam momentum of 3.01~GeV/c our Regge model as well 
as the GWU PWA describe the data. At the higher momentum
the GWU PWA already fails to reproduce the data, while our solution 
is still in rather good agreement with the experimental information. 

Fig.~\ref{compa3} shows the double polarization parameter $A_{NN}$ at
beam momenta of 3.01 and 3.511 GeV/c. The data are from measurements
at SATURNE~\cite{Allgower01,Lehar87}. 
Here the Regge approach agrees with both data and the GWU PWA for
$|t| < 0.5$ GeV$^2$ for the beam momentum 3.511 GeV/c. 

\section{Summary}
We performed a systematic analysis of $pp$ scattering at beam
momenta from 3 GeV/c up to 50 GeV/c 
utilizing the Regge formalism.
Experimental results on differential cross sections, analyzing powers, and
double polarization parameters available from recent measurements at 
COSY-EDDA, SATURNE as well as data collected previously at ZGS ANL allow to
construct the Regge amplitudes for $pp$ scattering with reasonable accuracy.
For momenta below 4 GeV/c 
there are no precise data at forward direction, i.e.
at scattering angles below 30$^\circ$ in the center-of-mass system.
Therefore, the amplitudes cannot be fixed in a unique way at these angles. 
Usually, the Regge
model works rather well at small four-momentum transfer squared or
forward angles and it would be crucial to collect new data in this
kinematical region. We expect that further progress in the analysis of
$pp$ scattering will become possible with forthcoming data from the 
ANKE COSY Collaboration~\cite{Chiladze09}, which will cover 
the momentum range analyzed here but will include angles in 
forward direction.

\begin{acknowledgement}
We are indebted to N.N. Nikolaev for instructive discussions. 
This work is partially supported by the Helmholtz Association through funds
provided to the virtual institute ``Spin and strong QCD'' (VH-VI-231), by
the EU Integrated Infrastructure Initiative  HadronPhysics2 Project (WP4
QCDnet) and by DFG (SFB/TR 16, ``Subnuclear Structure of Matter''). This
work was also supported in part by U.S. DOE Contract No. DE-AC05-06OR23177,
under which Jefferson Science Associates, LLC, operates Jefferson Lab. A.S.
acknowledges support by the JLab grant SURA-06-C0452 and the COSY FFE
grant No. 41760632 (COSY-085). 
\end{acknowledgement}

\section{Appendix}
In this appendix we summarize information on the proton-proton scattering
data analyzed in the present work.
\begin{table}[t]
\begin{center}
\caption{\label{difcros} References to data on differential cross 
sections for elastic $pp$ scattering analyzed in the present
work. Here $p$ is the proton beam momentum in the laboratory system, and
$t_{min}$ and $t_{max}$ denote the minimal and maximal four-momentum 
transfer squared, respectively.}
\renewcommand{\arraystretch}{1.1}
\begin{tabular}{|c|r|r|l|c|}
\hline
$p$ & $t_{min}$ & $t_{max}$ & Experiment & Ref.  \\
(GeV/c) & GeV$^2$ & GeV$^2$ & & \\
\hline
  3.012 & --2.043& --0.419  & COSY-EDDA & \cite{Albers04} \\
  3.017 & --2.075 & --1.225 & ZGS ANL  &  \cite{Jenkins78,Jenkins80} \\
  3.036 & --2.064& --0.423  & COSY-EDDA & \cite{Albers04} \\
  3.062 & --2.087& --0.428  & COSY-EDDA & \cite{Albers04} \\
  3.087 & --2.109 & --0.432 & COSY-EDDA & \cite{Albers04}  \\
  3.095 & --2.125 & --1.225 & ZGS ANL  &  \cite{Jenkins78,Jenkins80} \\
  3.112 & --2.131 & --0.437 & COSY-EDDA & \cite{Albers04} \\
  3.137 & --2.153 & --0.441 & COSY-EDDA & \cite{Albers04}  \\
  3.150 & --2.175 & --1.375 & ZGS ANL  &  \cite{Jenkins78,Jenkins80}  \\
  3.162 & --2.175 & --0.446 & COSY-EDDA & \cite{Albers04} \\
  3.187 & --2.197 & --0.450 & COSY-EDDA & \cite{Albers04}   \\
  3.205 & --2.225 & --1.325 & ZGS ANL  &  \cite{Jenkins78,Jenkins80} \\
  3.212 & --2.219 & --0.455 & COSY-EDDA & \cite{Albers04}  \\
  3.237 & --2.242 & --0.459 & COSY-EDDA & \cite{Albers04}  \\
  3.262 & --2.264 & --0.513 & COSY-EDDA & \cite{Albers04}   \\
  3.266 & --2.275 & --1.275 & ZGS ANL  &  \cite{Jenkins78,Jenkins80}  \\
  3.287 & --2.286 & --0.518 & COSY-EDDA & \cite{Albers04}   \\
  3.312 & --2.308 & --0.523 & COSY-EDDA & \cite{Albers04}  \\
  3.337 & --2.330 & --0.529 & COSY-EDDA & \cite{Albers04}   \\
  3.350 & --2.325 & --1.325 & ZGS ANL  &  \cite{Jenkins78,Jenkins80}  \\
  3.362 & --2.352 & --0.534 & COSY-EDDA & \cite{Albers04} \\
  3.388 & --2.375 & --0.539 & COSY-EDDA & \cite{Albers04}   \\
  3.410 & --2.350 & --1.300 & ZGS ANL  &  \cite{Jenkins78,Jenkins80} \\
  3.469 & --2.399 & --1.450 & ZGS ANL  &  \cite{Jenkins78,Jenkins80}   \\
  3.499 & --2.501 & --0.379 & ZGS ANL  & \cite{Kammerud71} \\
  3.530 & --2.525 & --1.375 & ZGS ANL  &  \cite{Jenkins78,Jenkins80}  \\
  3.621 & --2.625 & --1.475 & ZGS ANL  &  \cite{Jenkins78,Jenkins80}  \\
  3.686 & --2.675 & --1.575 & ZGS ANL  &  \cite{Jenkins78,Jenkins80}  \\
  3.750 & --2.725 & --1.575 & ZGS ANL  &  \cite{Jenkins78,Jenkins80}  \\
  3.843 & --2.850 & --1.350 & ZGS ANL  &  \cite{Jenkins78,Jenkins80}   \\
  3.942 & --2.850 & --1.450 & ZGS ANL  &  \cite{Jenkins78,Jenkins80}  \\
  4.013 & --2.850 & --1.550 & ZGS ANL  &  \cite{Jenkins78,Jenkins80}  \\
  4.082 & --2.950 & --1.750 & ZGS ANL  &  \cite{Jenkins78,Jenkins80}   \\
  4.151 & --3.050 & --1.650 & ZGS ANL  &  \cite{Jenkins78,Jenkins80}  \\
  4.2   & --0.0188 & --1.6${\cdot}10^{-3}$ &  CERN PS & \cite{Jenni77} \\ 
  4.258 & --3.150 & --1.750 & ZGS ANL  &  \cite{Jenkins78,Jenkins80}  \\
  4.334 & --3.150 & --1.850 & ZGS ANL  &  \cite{Jenkins78,Jenkins80}  \\
  4.409 & --3.250 & --1.950 & ZGS ANL  &  \cite{Jenkins78,Jenkins80}   \\
  4.483 & --3.250 & --1.650 & ZGS ANL  &  \cite{Jenkins78,Jenkins80}   \\
  4.559 & --3.450 & --1.750 & ZGS ANL  &  \cite{Jenkins78,Jenkins80}   \\
  4.681 & --3.550 & --1.950 & ZGS ANL  &  \cite{Jenkins78,Jenkins80}  \\
  6.800 & --0.863 & --0.023 & AGS BNL & \cite{Folley63}  \\
\hline
\end{tabular}
\end{center}
\end{table}

\begin{table}[t]
\begin{center}
\caption{\label{difcros1} References to data on differential cross 
sections for elastic $pp$ scattering analyzed in the present
work. 
Here $p$ is the proton beam momentum in the laboratory system,
and $t_{min}$ and $t_{max}$ denote the minimal and maximal four-momentum 
transfer squared, respectively.}
\renewcommand{\arraystretch}{1.1}
\begin{tabular}{|c|r|r|l|c|}
\hline
$p$ & $t_{min}$ & $t_{max}$ & Experiment & Ref.  \\
(GeV/c) & GeV$^2$ & GeV$^2$ & & \\
\hline
  7.0   & --5.85${\cdot}10^{-2}$   & --1.41${\cdot}10^{-3}$ &   CERN PS &
\cite{Jenni77} \\ 
  8.000 & --3.500 & --1.740 & AGS BNL   &  \cite{Orear66} \\
  8.500 & --1.050 & --0.130 & CERN PS  & \cite{Harting65}  \\
  8.800 & --0.916 & --0.039 &  AGS BNL  & \cite{Folley63}  \\
  10.0   & --9.72${\cdot}10^{-2}$   & --1.8${\cdot}10^{-3}$ &   CERN PS &
\cite{Jenni77} \\ 
  10.0 & --2.05 & --0.135 & CERN PS & \cite{Allaby73} \\
 10.800 & --0.824 & --0.058 & AGS BNL  & \cite{Folley63}   \\
 10.940 & --0.891 & --0.200 &  AGS BNL &  \cite{Folley65} \\
 12.000 & --2.500 & --1.200 & AGS BNL   & \cite{Orear66}  \\
  12.0 & --2.74 & --0.104 & CERN PS & \cite{Allaby73} \\
 12.100 & --0.342 & --0.011 & CERN PS  &  \cite{Diddens62} \\
 12.400 & --2.000 & --0.130 &  CERN PS  &   \cite{Harting65}  \\
 12.800 & --0.856 & --0.049 & AGS BNL   & \cite{Folley63}   \\
 13.16 & --0.104 & --9.2$\cdot$10$^{-3}$ & CERN PS   & \cite{Bez}   \\
  14.2 & --3.54 & --0.273 & CERN PS & \cite{Allaby73} \\
 14.800 & --0.781 & --0.066 &  AGS BNL  & \cite{Folley63}  \\
 14.930 & --0.736 & --0.216 &  AGS BNL &  \cite{Folley65} \\
 15.500 & --0.563 & --0.019 &  CERN PS  &  \cite{Diddens62} \\
 15.52 & --0.107 & --8.9$\cdot$10$^{-3}$ & CERN PS   & \cite{Bez}   \\
 16.700 & --0.698 & --0.042 & AGS BNL  & \cite{Folley63}  \\
 18.400 & --3.600 & --0.190 & CERN PS  & \cite{Harting65}  \\
 18.600 & --0.794 & --0.036 & CERN PS   &  \cite{Diddens62}  \\
 19.600 & --0.811 & --0.115 & AGS BNL  & \cite{Folley63}   \\
 19.840 & --0.787 & --0.230 &  AGS BNL &  \cite{Folley65} \\
 21.400 & --1.055 & --0.032 & CERN PS   &  \cite{Diddens62}  \\
 21.880 & --0.807 & --0.235 & AGS BNL  & \cite{Folley65}  \\
  24.0 & --6.72 & --0.0828 & CERN PS & \cite{Allaby73} \\
 24.56 & --0.111 & --1.1$\cdot$10$^{-2}$ & CERN PS   & \cite{Bez}   \\
 24.630 & --0.748 & --0.254 &  AGS BNL  &  \cite{Folley65} \\
 26.200 & --1.042 & --0.064 & CERN PS   & \cite{Diddens62}   \\
 30.00 & --0.119 & --2.7$\cdot$10$^{-3}$ & CERN PS   & \cite{Gesh}   \\
 30.45 & --0.113 & --1.1$\cdot$10$^{-2}$ & CERN PS   & \cite{Bez}   \\
 44.500 & --2.003 & --0.161 &  Serpukhov & \cite{Bruneton77}  \\
 45.17 & --0.115 & --1.1$\cdot$10$^{-2}$ & CERN PS   & \cite{Bez}   \\
 50.000 & --4.00 & --0.8 &  CERN SPS & \cite{Asad84} \\
\hline
\end{tabular}
\end{center}
\end{table}

\begin{table}[t]
\begin{center}
\caption{\label{apower} References to data on analyzing powers $A$ and
polarizations $P$ for elastic $pp$ scattering ($A{=}P$) 
analyzed in the present work. 
Here $p$ is the proton beam momentum in the laboratory system,
and $t_{min}$ and $t_{max}$ denote the minimal and maximal four-momentum 
transfer squared, respectively.
}
\renewcommand{\arraystretch}{1.1}
\begin{tabular}{|c|r|r|l|c|}
\hline
$p$ & $t_{min}$ & $t_{max}$ & Experiment & Ref.  \\
(GeV/c) & GeV$^2$ & GeV$^2$ & & \\
\hline
2.999 & --2.349& --0.988  & SATURNE & \cite{Allgover99} \\
3.00 & --2.803& --0.082  & ZGS ANL & \cite{Miller77} \\
3.01 & --2.359& --0.994  & SATURNE & \cite{Allgover99} \\
3.015 &  --2.009 & --0.316 &  COSY-EDDA & \cite{Altmeier00,Altmeier05} \\
3.021 & --2.37 & --1.0  & SATURNE & \cite{Allgover99} \\
3.031 & --2.38 & --1.05  & SATURNE & \cite{Allgover99} \\
3.045 & --2.035 &  --0.320 & COSY-EDDA & \cite{Altmeier00,Altmeier05} \\
3.075 & --2.061 & --0.325 & COSY-EDDA & \cite{Altmeier00,Altmeier05} \\
3.105 & --2.087 & --0.329 & COSY-EDDA & \cite{Altmeier00,Altmeier05} \\
3.135 & --2.113 & --0.333 & COSY-EDDA & \cite{Altmeier00,Altmeier05} \\
3.146 & --2.716 & --1.35  & SATURNE & \cite{Allgover99a} \\
3.165 & --2.139 & --0.337 & COSY-EDDA & \cite{Altmeier00,Altmeier05} \\
3.195 & --2.165 & --0.341 & COSY-EDDA & \cite{Altmeier00,Altmeier05} \\
3.196 & --2.4 & --0.156 & SATURNE & \cite{Perrot87} \\
3.198 & --2.778 & --1.129 & SATURNE & \cite{Allgover99a} \\
3.225 & --2.191 & --0.345 & COSY-EDDA & \cite{Altmeier00,Altmeier05} \\
3.25 & --2.604 & --0.232 & ZGS ANL & \cite{Parry73}  \\
3.251 & --2.844 & --1.137 & SATURNE & \cite{Allgover99a} \\
3.29 & --2.953 & --1.187 & SATURNE & \cite{Allgover99a} \\
3.3 & --2.257 & --0.355 & COSY-EDDA & \cite{Altmeier00,Altmeier05} \\
3.32 & --2.979 & --1.2 & SATURNE & \cite{Allgover99a} \\
3.37 & --3.043 & --1.228 & SATURNE & \cite{Allgover99a} \\
3.39 & --3.054 & --1.233 & SATURNE & \cite{Allgover99a} \\
3.41 & --3.081 & --1.243 & SATURNE & \cite{Allgover99a} \\
3.46 & --3.137 & --1.268 & SATURNE & \cite{Allgover99a} \\
3.508 & --3.052 & --0.129 & SATURNE & \cite{Perrot87} \\
3.61 & --3.263 & --1.428 & SATURNE & \cite{Allgover99a} \\
3.75 & --2.298 & --0.258 & ZGS ANL & \cite{Parry73}  \\
3.83 & --2.783 & --0.383 & SATURNE & \cite{Deregel76}  \\
4.00 & --2.989& --0.097  & ZGS ANL & \cite{Miller77} \\
4.40 & --2.96 & --0.233 & ZGS ANL & \cite{Parry73}  \\
5.15 & --4.055 & --0.341  & ZGS ANL & \cite{Parry73}  \\
5.15 & --4.031 & --0.54 &  ZGS ANL & \cite{Abshire74} \\
6.00 & --1.926& --0.098  & ZGS ANL & \cite{Miller77} \\
6.00 & --2.5& --0.05  & CERN PS & \cite{Borghini70} \\
6.00 & --4.62& --1.40  & ZGS ANL & \cite{Linn82} \\
7.0 & --5.245 & --0.582 & ZGS ANL & \cite{Abshire74} \\
10.0 & --2.9 & --0.08 & CERN PS & \cite{Borghini71} \\
11.75 & --2.509 & --0.618 & ZGS ANL & \cite{Abe76} \\
11.8 & --0.9 & --0.15 & ZGS ANL & \cite{Kramer71} \\
12.33 & --6.191 & --1.531 & ZGS ANL & \cite{Abshire74} \\
14.0 & --2.0 & --0.1 & CERN PS & \cite{Borghini71} \\
17.0 & --2.64 & --0.17 & CERN PS & \cite{Borghini71} \\
20.0 & --1.0 & --0.3 & FNAL &\cite{Corcoran80} \\
24.0 & --5.0 & --0.7 & CERN PS & \cite{Antille81} \\
28.0 & --2.997 & --0.525 & AGS BNL   & \cite{Hansen83} \\
45.0 & --1.0 & --0.3 & FNAL &\cite{Corcoran80} \\
\hline
\end{tabular}
\end{center}
\end{table}

\begin{table}[t]
\begin{center}
\caption{\label{ann} References to data on the double polarization 
parameters $A_{NN}$ and $C_{NN}$, ($A_{NN}{=}C_{NN}$) for 
elastic $pp$ scattering analyzed in the present work. 
Here $p$ is the proton beam momentum in the laboratory system,
and $t_{min}$ and $t_{max}$ denote the minimal and maximal four-momentum 
transfer squared, respectively.
}
\renewcommand{\arraystretch}{1.1}
\begin{tabular}{|c|r|r|l|c|}
\hline
$p$ & $t_{min}$ & $t_{max}$ & Experiment & Ref.  \\
(GeV/c) & GeV$^2$ & GeV$^2$ & & \\
\hline
  2.999 & --2.672 & --0.097 & ZGS ANL  &  \cite{Miller77}  \\
  2.999 & --2.348 & --0.987 & SATURNE  & \cite{Allgower00}  \\
  3.010 & --2.502 & --0.998 & SATURNE  & \cite{Allgower01}  \\
  3.010 & --2.360 & --0.992 & SATURNE  & \cite{Allgower00}  \\
  3.020 & --2.563 & --1.002 & SATURNE  & \cite{Allgower01} \\
  3.020 & --2.370 & --0.999 & SATURNE  & \cite{Allgower00} \\
  3.030 & --2.574 & --1.010 & SATURNE  & \cite{Allgower01}  \\
  3.030 & --2.376 & --1.051 & SATURNE  & \cite{Allgower00}  \\
  3.100 & --2.063 & --0.338 & COSY-EDDA  & \cite{Bauer05}  \\
  3.146 & --2.713 & --1.034 & SATURNE  & \cite{Allgower01}  \\
  3.180 & --2.133 & --0.349 & COSY-EDDA  & \cite{Bauer05} \\
  3.198 & --2.778 & --1.130 & SATURNE  & \cite{Allgower01}   \\
  3.199 & --2.513 & --0.120 & SATURNE  & \cite{Lehar87}  \\
  3.250 & --2.844 & --1.136 & SATURNE  & \cite{Allgower01}  \\
  3.300 & --2.235 & --0.366 & COSY-EDDA  & \cite{Bauer05}  \\
  3.302 & --2.954 & --1.188 & SATURNE  & \cite{Allgower01}  \\
  3.320 & --2.974 & --1.200 & SATURNE  & \cite{Allgower01}  \\
  3.370 & --3.035 & --1.226 & SATURNE  & \cite{Allgower01} \\
  3.3900 & --3.058 & --1.235 & SATURNE  & \cite{Allgower01}  \\
  3.4100 & --3.085 & --1.245 & SATURNE  & \cite{Allgower01}  \\
  3.4600 & --3.138 & --1.268 & SATURNE  & \cite{Allgower01}  \\
  3.5110 & --3.033 & --0.131 & SATURNE  & \cite{Lehar87}  \\
  3.6130 & --3.260 & --1.427 &  SATURNE  & \cite{Allgower01}  \\
  4.0 & --2.803 & --0.120 & ZGS ANL  &  \cite{Miller77}  \\
\hline
\end{tabular}
\end{center}
\end{table}


\end{document}